\documentclass{sigchi}

\CopyrightYear{2018}% 
\setcopyright{rightsretained}%
\conferenceinfo{UIST '18}{October 14--17, 2018, Berlin, Germany}%
\isbn{978-1-4503-5948-1/18/10}%
\doi{https://doi.org/10.1145/3242587.3242592}%

\usepackage{balance}       % to better equalize the last page
\usepackage{graphics}      % for EPS, load graphicx instead 
\usepackage[T1]{fontenc}   % for umlauts and other diaeresis
\usepackage{txfonts}
\usepackage{mathptmx}
\usepackage[pdflang={en-US},pdftex]{hyperref}
\usepackage{color}
\usepackage{booktabs}
\usepackage{textcomp}

\usepackage{microtype}        % Improved Tracking and Kerning
\usepackage{ccicons}          % Cite your images correctly!
\usepackage{todonotes}

\usepackage[]{algorithm2e}
\usepackage{amssymb}
\usepackage{amsmath}
\usepackage{graphicx}
\usepackage{multirow}
\usepackage{array}
\usepackage{todonotes}
\usepackage{bbold}
\usepackage{booktabs}
\usepackage{balance}

\newcommand{\fig}[1]{Figure~\ref{#1}}

\newcommand{\tbl}[1]{Table~\ref{#1}}

\newcommand{\myparagraph}[1]{\vspace{.5\baselineskip}\noindent{\bf #1}}
\newcommand{\myitem}{\vspace{-5pt}\item}

\newcommand{\change}[1]{{\color{black} #1}}

\def\sq{\hspace{-2.5pt}}
\def\sq2{\hspace{-2pt}}

\def\modelname{MoSculp\xspace}

\definecolor{MyDarkBlue}{rgb}{0,0.08,0.5}
\definecolor{MyDarkGreen}{rgb}{0,0.70,1}
\definecolor{MyDarkRed}{rgb}{0.7,0,0}
\definecolor{MyDarkOrange}{rgb}{0.40,0.2,0}
\definecolor{MyPurple}{RGB}{111,0,255}
\definecolor{MyRed}{rgb}{1.0,0.0,0.0}

\newcommand{\ignore}[1]{} % <- Use to comments a large chunk of text

\newcommand{\tempComment}[1]{}

\makeatletter
\DeclareRobustCommand\onedot{\futurelet\@let@token\@onedot}
\def\@onedot{\ifx\@let@token.\else.\null\fi\xspace}

\def\eg{\emph{e.g}\onedot} 
\def\ie{\emph{i.e}\onedot} 
 
\def\etc{\emph{etc}\onedot} \def\vs{\emph{vs}\onedot}
 
\def\etal{\emph{et al}\onedot}
\makeatother

\def\summarynames{{motion sculptures}\xspace} % shape-time action summary
\def\summaryname{{motion sculpture}\xspace} % shape-time action summary
\def\Summarynames{{Motion sculptures}\xspace}
\def\SummaryNames{{Motion Sculptures}\xspace}
\def\Summaryname{{Motion sculpture}\xspace}
\def\SummaryName{{Motion Sculpture}\xspace}

\def\plaintitle{SIGCHI Conference Proceedings Format}
\def\plainauthor{First Author, Second Author, Third Author,
  Fourth Author, Fifth Author, Sixth Author}

\makeatletter
\def\url@leostyle{%
  \@ifundefined{selectfont}{
    \def\UrlFont{\sf}
  }{
    \def\UrlFont{\small\bf\ttfamily}
  }}
\makeatother
\def\pprw{8.5in}
\def\pprh{11in}

\definecolor{linkColor}{RGB}{6,125,233}
\hypersetup{%
  pdftitle={\plaintitle},
  pdfauthor={\plainauthor},
  pdfkeywords={Motion visualization; motion sculptures},
  pdfdisplaydoctitle=true, % For Accessibility
  bookmarksnumbered,
  pdfstartview={FitH},
  colorlinks,
  citecolor=black,
  filecolor=black,
  linkcolor=black,
  urlcolor=linkColor,
  breaklinks=true,
  hypertexnames=false
}

\title{\modelname: Interactive Visualization of Shape and Time}
\author{%
  Xiuming Zhang $^1$\qquad
  Tali Dekel $^2$\qquad
  Tianfan Xue $^2$\qquad
  Andrew Owens $^3$\qquad
  Qiurui He $^1$\vspace{.2em}
  \\
  Jiajun Wu $^1$\qquad
  Stefanie Mueller $^1$\qquad
  William T.\ Freeman $^{1,2}$\vspace{.3em}
  \\
  \affaddr{$^1$ MIT CSAIL}\hspace{6em}
  \affaddr{$^2$ Google Research}\hspace{6em}
  \affaddr{$^3$ UC Berkeley}\vspace{.2em}
  \\
  \email{\normalsize\{xiuming, q\_he, jiajunwu, stefanie.mueller, billf\}@mit.edu}\hspace{2em}
  \email{\normalsize\{tdekel, tianfan\}@google.com}\hspace{2em}
  \email{\normalsize owens@berkeley.edu}
}

\begin{document}

\maketitle

\begin{abstract}

We present a system that visualizes complex human motion via 3D \emph{\summarynames}---a representation that conveys the 3D structure swept by a human body as it moves through space. Our system computes a \summaryname from an input video, and then embeds it back into the scene in a 3D-aware fashion. The user may also  explore the sculpture directly in 3D or physically print it. Our interactive interface allows users to customize the sculpture design, for example, by selecting materials and lighting conditions.
%or rendering it in a synthetic scene.

To provide this end-to-end workflow, we introduce an algorithm that estimates a human's 3D geometry over time from a set of 2D images, and develop a 3D-aware image-based rendering approach that inserts the sculpture back into the original video. By automating the process, our system takes \summaryname creation out of the realm of professional artists, and makes it applicable to a wide range of existing video material.

By conveying 3D information to users, \summarynames reveal space-time motion information that is difficult to perceive with the naked eye, and allow viewers to interpret how different parts of the object interact over time. We validate the effectiveness of \summarynames with user studies, finding that our visualizations are more informative about motion than existing stroboscopic and space-time visualization methods.

\end{abstract}

\begin{CCSXML}
<ccs2012>
<concept>
<concept_id>10003120.10003145.10003147.10010923</concept_id>
<concept_desc>Human-centered computing~Information visualization</concept_desc>
<concept_significance>500</concept_significance>
</concept>
</ccs2012>
\end{CCSXML}
\ccsdesc[500]{ Human-centered computing~Information visualization}
\printccsdesc
%\category{H.5.2.}{Information Interfaces and Presentation: User Interfaces.}{}{}

\keywords{Motion estimation and visualization; image-based rendering}

\section{Introduction}
\label{sec:intro}

\begin{figure}[t]
 	\includegraphics[width=\columnwidth]{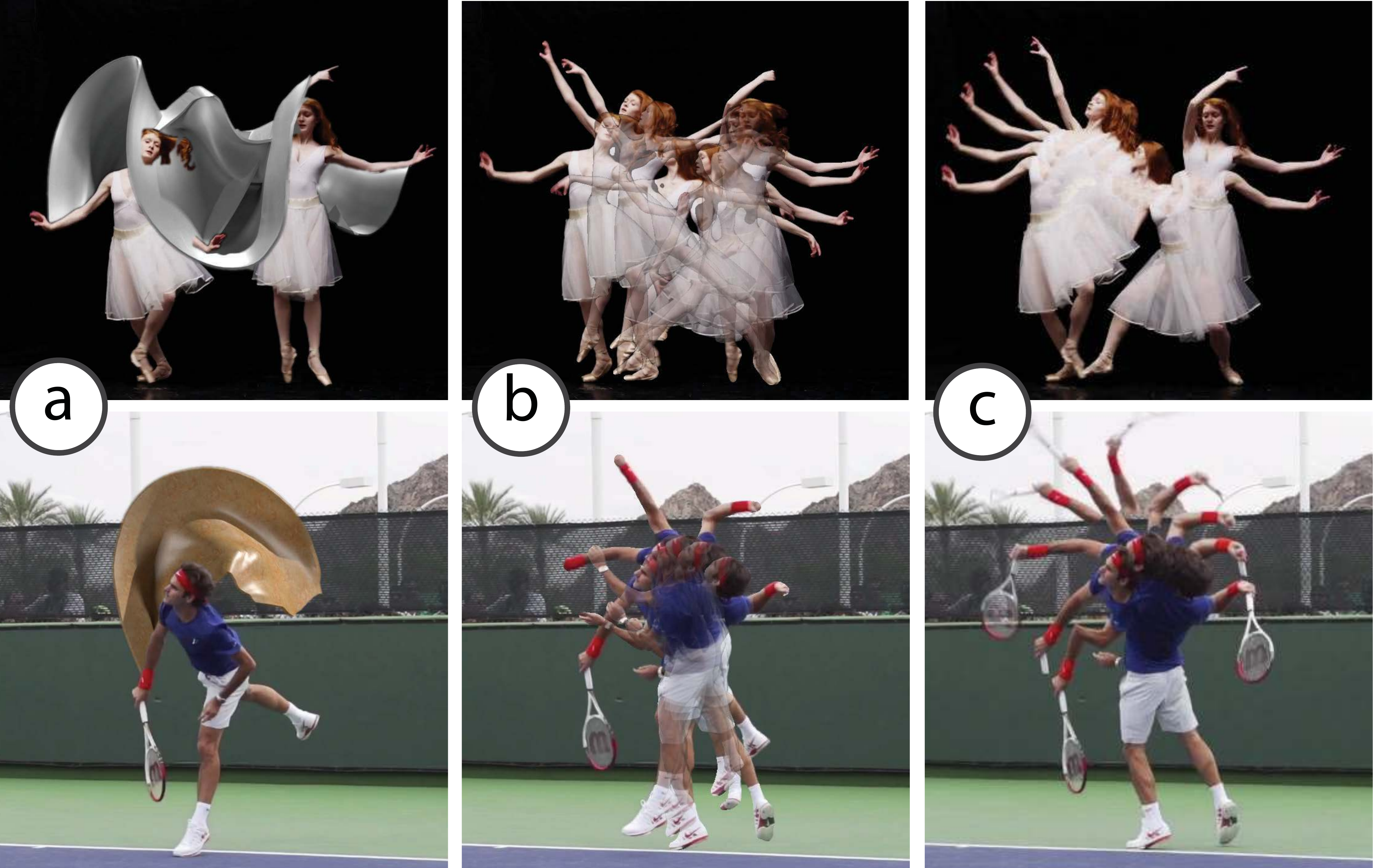}
 	%\vspace{-15pt}
	\caption{Comparison of (a) our \summarynames with (b)  stroboscopic photography and (c) shape-time photography~\protect\cite{freeman:shapetime} on the \textit{Ballet-1}~\protect\cite{ballet} and \textit{Federer} clips.
	%, which preserves the depth ordering of the collection of human shapes.
	}
 	\vspace{-5pt}
	\label{fig:comparison}
 \end{figure}

% motivation, problem
Complicated actions, such as swinging a tennis racket or dancing ballet, can be difficult to convey to a viewer through a static photo. To address this problem, researchers and artists have developed a number of motion visualization techniques, such as chronophotography, stroboscopic photography, and multi-exposure photography~\cite{muybridge1985horses,braun1992picturing}. However, since such methods operate entirely in 2D, they are unable to convey the motion's underlying 3D structure. Consequently, they tend to generate cluttered results when parts of the object are occluded (\fig{fig:comparison}). Moreover, they often require special capturing procedures, environment (such as a clean, black background), or lighting equipment.  

%such as chronophotography, stroboscopic photography, and multi-exposure photography~\cite{muybridge1985horses,braun1992picturing}

% \teaser{
% \includegraphics[width=\textwidth]{fig/teaser.pdf}
% \caption{We present a system for visualizing motion in a video via  \emph{\summaryname}---an artistic rendering of the 3D path an object traces as it moves through space. Our system transforms (a) a standard RGB video that depicts a complex action (captured by a static or moving camera) into either (b) a physical motion sculpture or (c) a rendering on top of the source video contents. The 3D nature of \summarynames reveals information about the motion, such as the sinusoidal motion of the arms, which is not readily visible in the input video. (d) In our user interface, users can provide input about the number of stick figures, materials, lighting, and scene background to use to further facilitate the understanding of the motion. }
% \label{fig:teaser}
% }

 \begin{figure}[t!]
 	\includegraphics[width=1\columnwidth]{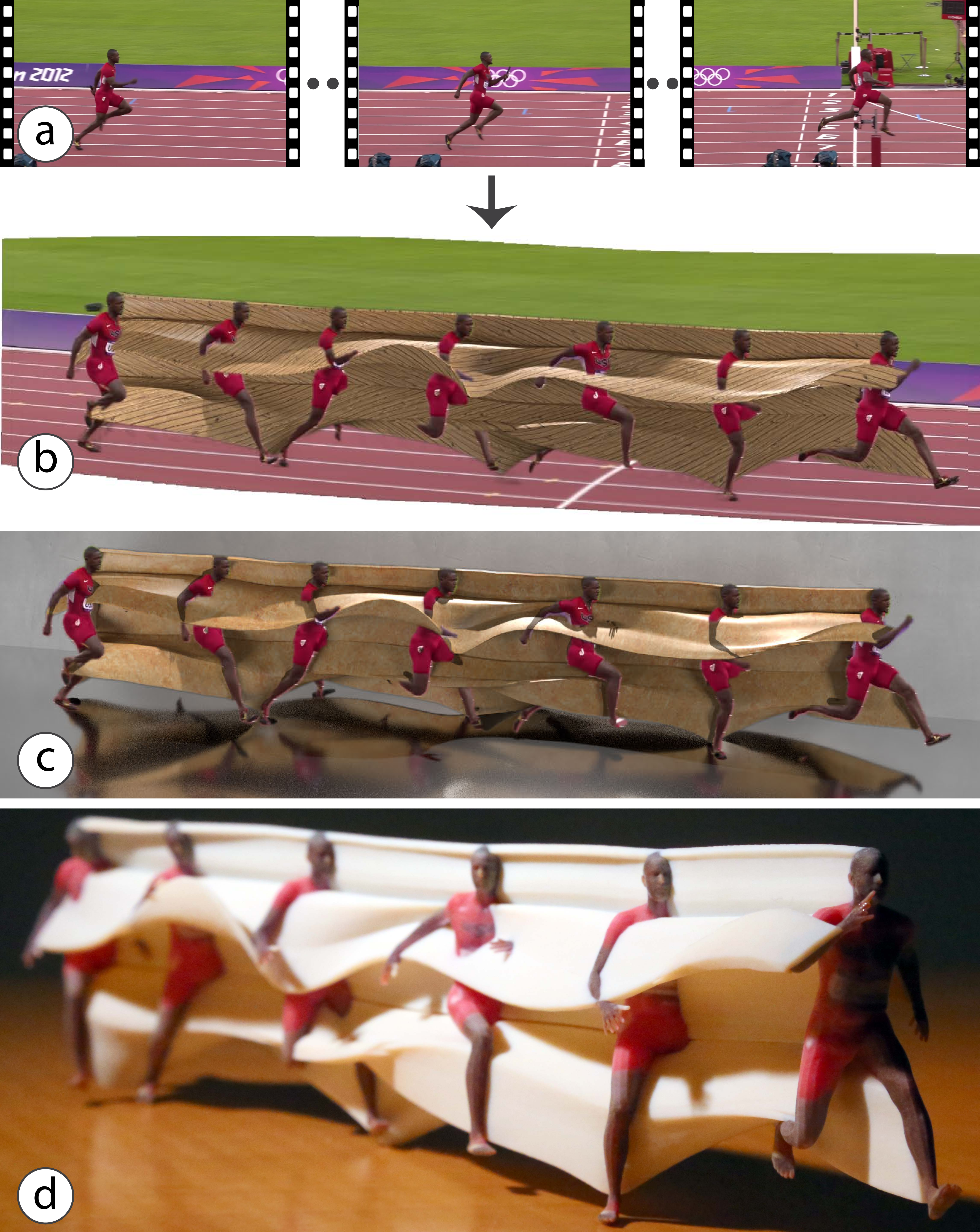}
	\caption{Our \modelname system transforms a video (a) into a \summaryname, \ie, the 3D path traced by the human while moving through space. Our \summarynames can be virtually inserted back into the original video (b), rendered in a synthetic scene (c), and physically 3D printed (d). Users can interactively customize their design, \eg, by changing the sculpture material and lighting.}
 	\vspace{-5pt}
	\label{fig:teaser}
 \end{figure}

In this paper, we present \emph{\modelname}, an end-to-end system that takes a video as input and produces a \emph{\summaryname}: a visualization of the spatiotemporal structure carved by a body as it moves through space. \Summarynames aid in visualizing the trajectory of the human body, and reveal how its 3D shape evolves over time. Once computed, \summarynames can be inserted back to the source video (\fig{fig:teaser}b), rendered in a synthesized scene (\fig{fig:teaser}c), or physically 3D printed (\fig{fig:teaser}d). 
%, making it easier the to understand human motion.
%In this paper, we present \emph{\modelname}, a system that facilitates the understanding of human motion captured in a standard video. Given a video, our system produces a \emph{\summaryname} -- visualization of the spatial-temporal structure carved by the body as it moves through space. \Summarynames convey a subject's trajectory and reveal how its 3D shape evolves over time. Once computed, \summarynames can be inserted back to the source video (\fig{fig:teaser}b), rendered in a synthesized scene (\fig{fig:teaser}c), or physically 3D-printed (\fig{fig:teaser}d). {\color{red} Our code is publicly available.}

% UI part
We develop an interactive interface that allows users to: (i) explore \summarynames in 3D, \ie, navigate around them and view them from alternative viewpoints, thus revealing information about the motion that is inaccessible from the original viewpoint, and (ii) customize various rendering settings, including lighting, sculpture material, body parts to render, scene background, and \etc\footnote{\change{Demo available at \url{http://mosculp.csail.mit.edu}}}. These tools provide flexibility for users to express their artistic designs, and further facilitate their understanding of human shape and motion.
%\mytodo{tali: not sure our experiments support this. We haven't evaluated the UI part.}
%\ao{Can you say more about these experiments here?}.

% contribution statement
 Our main contribution is devising the first end-to-end system for creating \summarynames from videos, thus making them accessible for novice users. A core component of our system is a method for estimating the human's pose and body shape over time. Our 3D estimation algorithm, built upon state of the art, has been designed to recover the 3D information required for constructing \summarynames (\eg, by modeling clothing), and to support simple user corrections. The \summaryname is then inferred from the union of the 3D shape estimations over time. To insert the sculpture back into the original video, we develop a 3D-aware, image-based rendering approach that preserves depth ordering. Our system achieves high-quality, artifact-free composites for a variety of human actions, such as ballet dancing, fencing, and other athletic actions.

% AO: Should we talk more concretely about the system and how the user interaction works (a) for making motion sculptures (b) for visualizing them? I'm worried that it sounds too much like the SIGGRAPH paper now.

% \begin{figure}[t]
%     \centering
%     %\begin{subfigure}{\linewidth}
%     \includegraphics[width=\linewidth]{fig/results_compare.pdf}
%     %\vspace{-15pt}
%     %\caption{We compare our \summaryname (a) with: shape-time photography~\cite{freeman:shapetime} (b), a depth-preserving 2.5D method, and the standard, depth-ignorant stroboscope (c).}
%     % \label{fig:comparison}
%     %\end{subfigure}

%     %\begin{subfigure}{\linewidth}
%     %\centering
%     %\vspace{15pt}
%     \setlength{\tabcolsep}{1.8pt}
%     \begin{tabular}{lcccccccccc}
%     \toprule
%     & Tenn & Bal1 & Bal2 & Jump & Walk & Olym & Dunk & {\bf Avg} \\
%     \midrule
%     Prefer Ours to \cite{freeman:shapetime} & 82 & 78 & 84 & 80 & 71 & 74 & 78 & {\bf 78} \\
%     \bottomrule
%     \end{tabular}
%     %\caption{On average, 78\% of the users agree that our visualization is better at conveying 3D motion than shape-time photography.}
%     %\label{tbl:comp-tbl}
%     %\end{subfigure}
%     \vspace{-0.3cm}
%     \caption{We compare with two summarization methods: (a) the standard, depth-ignorant stroboscope, and (b) shape-time photography~\protect\cite{freeman:shapetime}. We have also conducted a human study to compare our visualization with~\protect\cite{freeman:shapetime}; most users suggested ours conveys more 3D information.}
%     \label{fig:comparison}
%     \vspace{-15pt}
% \end{figure}
\section{Related Work}
% motivation, problem
We briefly review related work in the areas of artistic rendering, motion effects in images, human pose estimation, video editing and summarization methods, and physical visualizations. 

\myparagraph{Automating Artistic Renderings.}
A range of tools have been developed to aid users in creating artist-inspired motion visualizations~\cite{cutting2002representing, authoringhumanmovements, schmid2010programmable, bouvier2007motion}. DemoDraw~\cite{authoringhumanmovements} allows users to generate drawing animations by physically acting out an action, motion capturing them, and then applying different stylizing filters. 

%Within HCI, it is well known that different types of visual representations convey information better than others [13]
Our work continues along this line of work and is inspired by artistic work that visualizes 3D shape and motion~\cite{jldesign2013,peterJ,gever2014,gremmler2016, chronofab}. However, these renderings are produced by professional artists and require special recording procedures or advanced computer graphics skills. In this paper, we opt to lower the barrier to entry and make the production of \summarynames less costly and more accessible for novice users. 

%AO: agree with Tali that we should remove this
%\tali{remove?}In addition, since our system is fully automated and thus obtains precise measurements of the shape and motion rather than the subjective impressions of an artist, we can thoroughly evaluate it against existing methods and produce a list of challenges when creating motion sculptures from a wide variety of video material. 

The most closely related work to ours in this category is ChronoFab~\cite{chronofab}, a system for creating \summarynames from 3D animations. However, a key difference is that ChronoFab requires a full 3D model of the object and its motion as input, which limits the practical use of ChronoFab, while our system directly takes a video as input and estimates the 3D shape and motion as part of the pipeline. %, like our system.  
% AO: not sure what "user-designed" motions here means
% DemoCut ~\cite{democut},

\myparagraph{Motion Effects in Static Images.}
Illustrating motion in a single image dates back to stroboscopic photography \cite{muybridge1985horses} and classical methods that design and add motion effects to an image (\eg, speedlines~\cite{masuch1999r}, motion tails~\cite{bennett2007computational, teramoto2010interactive}, and motion blur).  Cutting~\cite{cutting2002representing} presented an interesting psychological standpoint and evaluation on the efficacy of different motion visualizations. In the context of non-photorealistic rendering, various motion effects have been designed for animations and cartoons~\cite{lake2000stylized, kawagishi2003cartoon}. Schmid~\etal~\cite{schmid2010programmable} designed programmable motion effects as part of a rendering pipeline %by aggregating triangle meshes over time and using this data structure 
to produce stylized blurring and stroboscopic images. Similar effects have also been produced by Baudisch~\etal~\cite{baudisch2006phosphor} for creating animated icon movements.  In comparison, our system does not require a 3D model of the object, but rather estimates it from a set of 2D images. In addition, most of these motion effects do not explicitly model the 3D aspects of motion and shape, which are the essence of \summarynames.  

%For time-lapse videos, Bennett~\etal~\cite{bennett2007computational} developed a method for simulating a virtual shutter that adds motion trails into the frames. 
% \change{Closely related to \summarynames is~\cite{blank2005actions}, which represents human actions as space-time shapes to improve action classification and clustering. As the space-time shapes are essentially 2D silhouettes protruding along time, later poses appear in front of (thus may occlude) earlier poses, despite that the earlier poses may be actually closer to the camera. \Summarynames, in contrast, make occlusion entirely geometric and time-independent by estimating 3D poses. }

\myparagraph{Video Editing and Summarization.}
\Summarynames are related to video \change{editing techniques, such as MovieReshape~\cite{Jain:2010:MovieReshape}, which  manipulates certain properties of the human body in a video,} and summarization techniques, such as image montage~\cite{agarwala2004interactive,sunkavalli2012video} that re-renders video contents in a more concise view, typically by stitching together foreground objects captured at different timestamps. As in stroboscopic photography, such methods do not preserve the actual depth ordering among objects, and thus cannot illustrate 3D information about shape and motion. Another related work is~\cite{blank2005actions}, which represents human actions as space-time shapes to improve action classification and clustering. However, their space-time shapes are 2D human silhouettes and thus do not convey 3D information. Video Summagator~\cite{nguyen2012video} visualizes a video as a space-time cube using volume rendering techniques. However, this approach does not model self-occlusions, which leads to clutter and visual artifacts.
% AO: not quite sure I understand the last sentence. Can we just say it doesn't handle self-occlusions, which leads to clutter and visuali artifacts during the volume inteprolation step?

% \input{figText/sculp_art.tex}

Depth-based summarization methods overcome some of these limitations using geometric information provided by depth sensors. Shape-time photography~\cite{freeman:shapetime}, for example, conveys occlusion relationships by showing, at each pixel, the color of the surface that is the closest to the camera over the entire video sequence. More recently, Klose~\etal introduced a video processing method that uses per-pixel depth layering to create action shot summaries~\cite{klose2015sampling}. While these methods are useful for presenting 3D relationships in a small number of sparsely sampled images, such as where the object is throughout the video, they are not well suited for visualizing continuous motion. Moreover, these methods are based on depth maps, and thus provide only a ``2.5D'' reconstruction that cannot be easily viewed from multiple viewpoints as in our case. 

 \begin{figure*}[t!]
 	\includegraphics[width=1\textwidth]{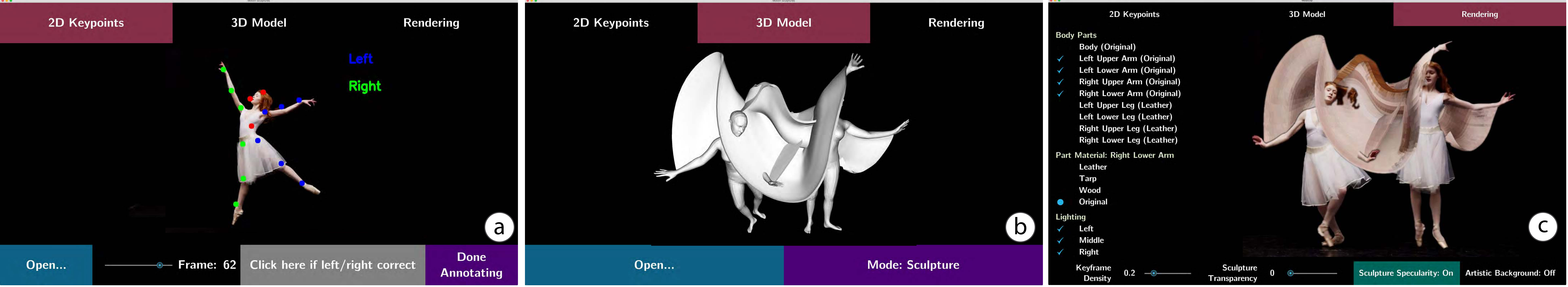}
	\caption{\modelname user interface. (a) The user can browse through the video and click on a few frames, in which the keypoints are all correct; these labeled frames are used to fix keypoint detection errors by temporal propagation.  After generating the \summaryname, the user can (b) navigate around it in 3D, and (c) customize the rendering by selecting which body parts form the sculpture, their materials, lighting settings, 
	%number of stick figures, i.e., human's snapshots that are embedded into the sculpture
	keyframe density, sculpture transparency, specularity, and the scene background.}
	\label{fig:userinterface}
 \end{figure*}

\myparagraph{Human Pose Estimation.}
\change{\Summaryname creation involves estimating the 3D human pose and shape over time -- a fundamental problem that has been extensively studied. Various methods have been proposed to estimate 3D pose from a single image \cite{bogo2016keep,kanazawa2017end,pavlakos2017coarse,pavlakos2018learning,pavlakos2018ordinal,chen20173d,tome2017lifting}, or from a video \cite{howe2000bayesian,huang2017towards,zhou2016sparseness,mehta2017vnect,agarwal2006recovering}. However, these methods are not designed for the specifics of motion visualization like our approach.}

\myparagraph{Physical Visualizations.}
Recent research has shown great progress in physical visualizations and demonstrated the benefit of allowing users to efficiently access information along all dimensions~\cite{physicalviz, tastybeats, proxyprint}. MakerVis~\cite{makervis} is a tool that allows users to quickly convert their digital information into physical visualizations. ChronoFab~\cite{chronofab}, in contrast, addresses some of the challenges in rendering digital data physical, \eg, connecting parts that would otherwise float in midair. Our \summarynames can be physically printed as well. However, our focus is in rendering and seamlessly compositing them into the source videos, rather than optimizing the procedure for physically printing them.

%Automatic generation of \summarynames poses two major challenges. First, we need to solve the inverse 2D-3D problem, \ie, recover the object's shape, pose, and motion from the source video. Second, given the estimated \summaryname, we want to blend it with image content in a visually pleasing way. This requires careful analysis because every little error in the sculpture may show up as visual artifacts in the final rendering. We address these challenges in the following manner.  

%In this paper, we focus on human motion, yet show that our approach can be extended to general objects for which can be reliably represented by a parametric 3D model. We formulate the problem of recovering the shape, pose, and trajectory of the object over time as a joint optimization problem. We utilize the temporal coherency of the human motion and body shape to constraint the problem and reduce ambiguities. With this estimations in hand, we form an initial motion sculpture. To obtain high quality, artifact-free result we developed an image-based rendering (IBR) that refines the sculpture and inserts it into a source video, while preserving the proper depth orderings. 

%Our generated sculptures on diverse videos including sport actions and dance, gracefully reveal the beauty and vividness in human motion. 
\section{System Walkthrough}

% Our system transforms a standard RGB video into a 3D motion sculpture, representing the 3D structure the human carves as it moves in space. The core part of our algorithm analyzes the video frames and estimates the body's 3D shape and pose over time. These  geometries are then swept to create the motion sculpture. 

% Once the processing step completes, the motion sculpture can be either physically printed (\fig{fig:teaser}d), or digitally rendered (\fig{fig:teaser}b-c). To this end, we developed an interactive system that allows users to load the motion sculpture, virtually explore it in 3D and customize its design by controlling various rendering settings. Our system's UI consists of three main windows shown in \fig{fig:userinterface}: 

% We developed an interactive system that allows users to load the motion sculpture, virtually explore it in 3D and customize its design by controlling various appearance settings. Our system's UI consists of three main windows as shown in \fig{fig:userinterface}.

% he 'generate motion sculpture' button, which triggers the system to estimate the object's shape and pose in all the frames. The initial step of this computation is to estimate the human's keypoints in each frame. For frames, in which the system has low confidence about which body part it identified (e.g. left vs. right arm), it queries information from the user via a simple either-or question. After the system finished computing all the frames, it renders the final 3D motion sculpture in the user interface.    

% overview of the approach 
To generate a \summaryname, the user starts by loading a video into the system, after which \modelname detects the 2D keypoints and overlays them on the input frames (\fig{fig:userinterface}a). The user then browses the detection results and confirms, on a few ($\sim$3-4) randomly selected frames, that the keypoints are correct by clicking the ``Left/Right Correct'' button. After labeling, the user hits ``Done Annotating,'' which triggers \modelname to correct temporally inconsistent detections, with these labeled frames serving as anchors. \modelname then generates the \summaryname in an offline process
%{\color{red}, taking 2-3 minutes per frame,}
that includes estimating the human's shape and pose in all the frames and rendering the sculpture.

After processing, the generated sculpture is loaded into \modelname, and the user can virtually explore it in 3D (\fig{fig:userinterface}b). This often reveals information about shape and motion that is not available from the original camera viewpoint, and facilities the understanding of how different body parts interact over time.

Finally, the rendered \summaryname is displayed in a new window (\fig{fig:userinterface}c), where the user can customize the design by controlling the following rendering settings.
 \begin{itemize}
     \myitem \myparagraph{Scene.} The user chooses to render the sculpture in a synthesized scene or embed it back into the original video by toggling the ``Artistic Background'' button in \fig{fig:userinterface}c. For synthetic scenes (\ie, ``Artistic Background'' on), we use a glossy floor and a simple wall lightly textured for realism. To help the viewer better perceive shape, we render shadows cast by the person and sculpture on the wall as well as their reflections on the floor (as can be seen in \fig{fig:teaser}c). 
     \myitem \myparagraph{Lighting.} Our set of lights includes two area lights on the left and right sides of the scene as well as a point light on the top. The user may choose any combination of these lights (see the ``Lighting'' menu in \fig{fig:userinterface}c).
     \myitem \change{\myparagraph{Body Parts.} The user decides which parts of the body form the \summaryname. For instance, one may choose to render only the arms to perceive clearly the arm movement, as in \fig{fig:comparison}a. The body parts that we consider are listed under the ``Body Parts'' menu in \fig{fig:userinterface}c.}
     \myitem \change{\myparagraph{Materials.} Users can control the texture of the sculpture by choosing one of the four different materials: leather, tarp, wood, and original texture (\ie, colors taken from the source video by simple ray casting). To better differentiate sculptures formed by different body parts, one can specify a different material for each body part (see the dynamically updating ``Part Material'' menu in \fig{fig:userinterface}c).}
     %\mytodo{I  prefer having Materials and  Body Parts as a different paragraphs. We can first discuss Materials and then just briefly mention that the user can assign each body part a different texture.}
     \myitem \myparagraph{Transparency.} A slider controls transparency of the \summaryname, allowing the viewer to see through the sculpture and better comprehend the complex space-time occlusion.
     \myitem \myparagraph{Human Figures.} In addition to the \summaryname, \modelname can also include a number of human images (similar to sparse stroboscopic photos), which allows the viewer to associate sculptures with the corresponding body parts that generated them. A density slider controls how many of these human images, sampled uniformly, get inserted.
\end{itemize} 

These tools grant users the ability to customize their visualization and select the rendering settings that best convey the space-time information captured by the \summaryname at hand.

\subsection{Example \SummaryNames}

We tested our system on a wide range of videos of complex actions including ballet, tennis, running, and fencing. We collected most of the videos from the Web (YouTube, Vimeo, and Adobe Stock), and captured two videos ourselves using a Canon 6D (\emph{Jumping} and \emph{U-Walking}). 

\begin{figure}[t!]
 	\includegraphics[width=\columnwidth]{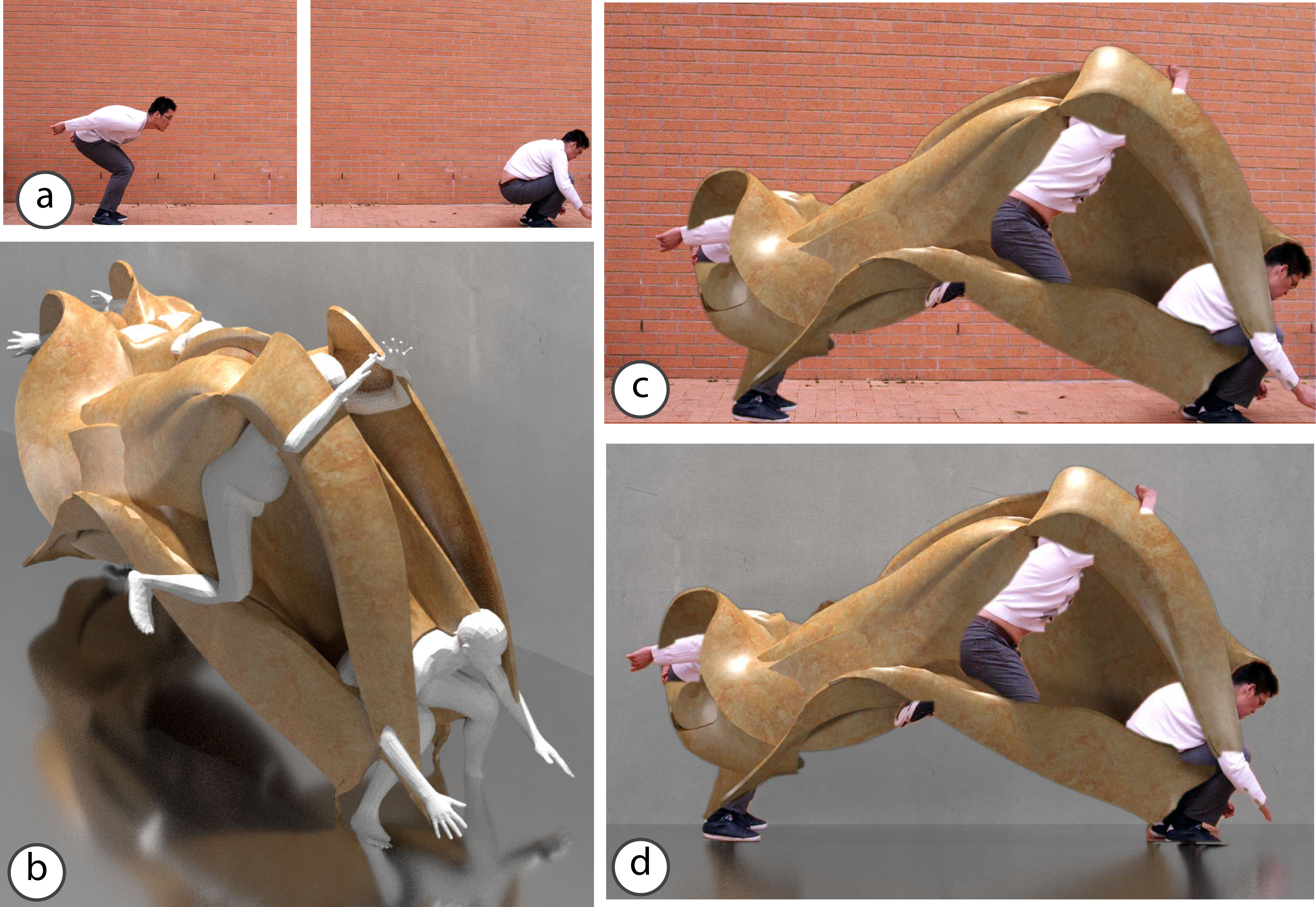}
 	%\vspace{-15pt}
	\caption{The \protect\textit{Jumping} sculpture (material: marble; rendered body parts: all). (a) First and final video frames. (b) Novel-view rendering. (c, d) The \summaryname is inserted back into the original scene and to a synthetic scene, respectively.}
 	\vspace{-5pt}
    \label{fig:jumping}
\end{figure}

For each example, we embed the \summaryname back into the source video and into a synthetic background. We also render the sculpture from novel viewpoints, which often reveals information imperceptible from the captured viewpoint. In \emph{Jumping} (\fig{fig:jumping}), for example, the novel-view rendering (\fig{fig:jumping}b) shows the slide-like structure carved out by the arms during the jump. 

\begin{figure}[t]
 	\includegraphics[width=\columnwidth]{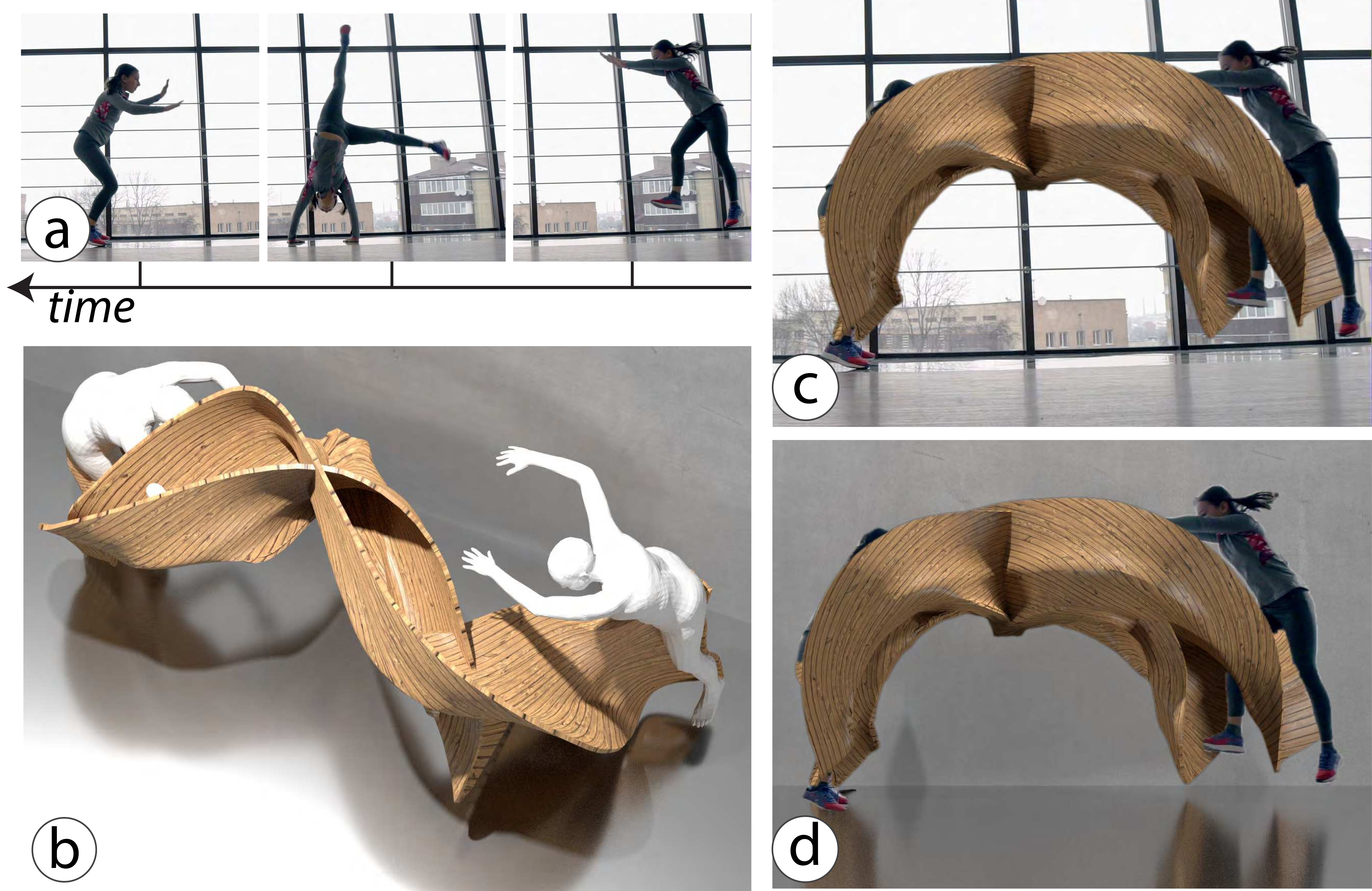}  
 	%\vspace{-5pt}
	\caption{The \textit{Cartwheel} sculpture (material: wood; rendered body parts: legs). (a) Sampled video frames. (b) Novel-view rendering. (c, d) The \summaryname is inserted back into the source video and to a synthetic scene, respectively.} %\tali{Do we want a detailed caption here? Given we already have defined the components in the jumping figure.} }
 	%\vspace{-5pt}
	\label{fig:cartwheel}
 \end{figure}

An even more complex action, cartwheel, is presented in \fig{fig:cartwheel}. For this example, we make use of the ``Body Parts'' options in our user interface, and decide to visualize only the legs to avoid clutter. Viewing the sculpture from a top view (\fig{fig:cartwheel}b) reveals that the girl's legs cross and switch their depth ordering---a complex interaction that is hard to comprehend even by repeatedly playing the original video.

\begin{figure}[t!]
 	\includegraphics[width=\columnwidth]{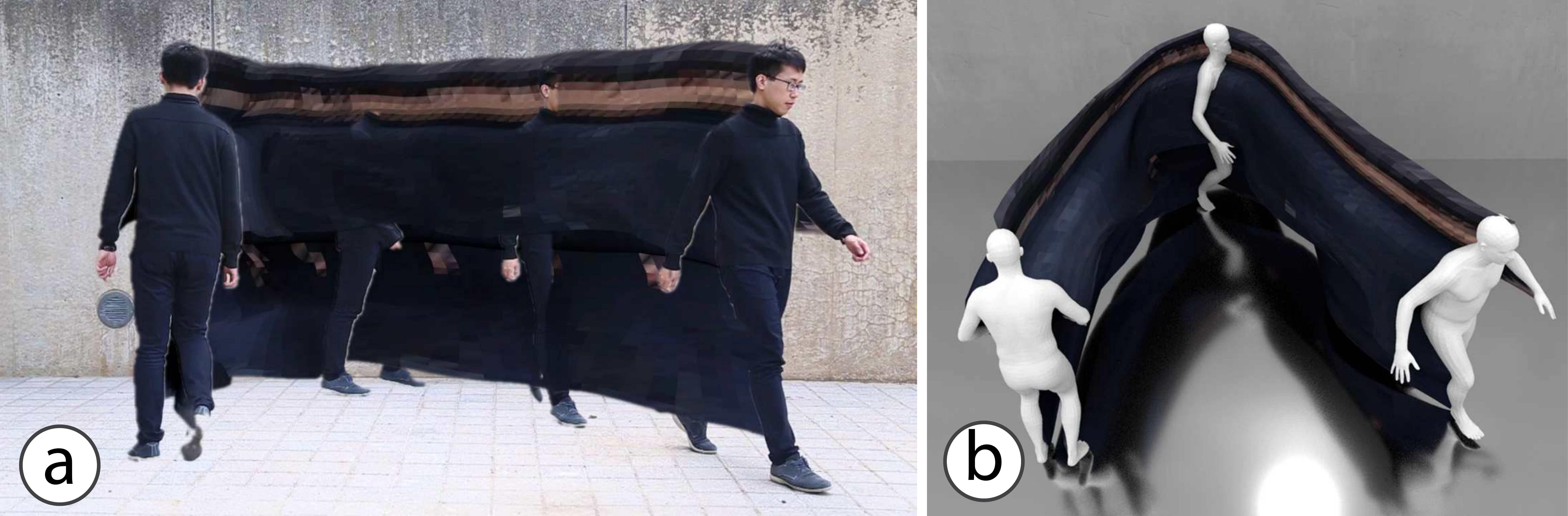}
 	%\vspace{-5pt}
	\caption{(a) The \textit{U-Walking} sculpture with texture taken from the source video. (b) The same sculpture rendered from a novel top view in 3D, which reveals the motion in depth.}
	\label{fig:walking}
 \end{figure}

In \emph{U-Walking} (\fig{fig:walking}), the \summaryname depicts the person's motion in depth; this can be perceived also from the original viewpoint (\fig{fig:walking}a), thanks to the shading and lighting effects that we select from the different rendering options.

\begin{figure}[t]
 	\includegraphics[width=\columnwidth]{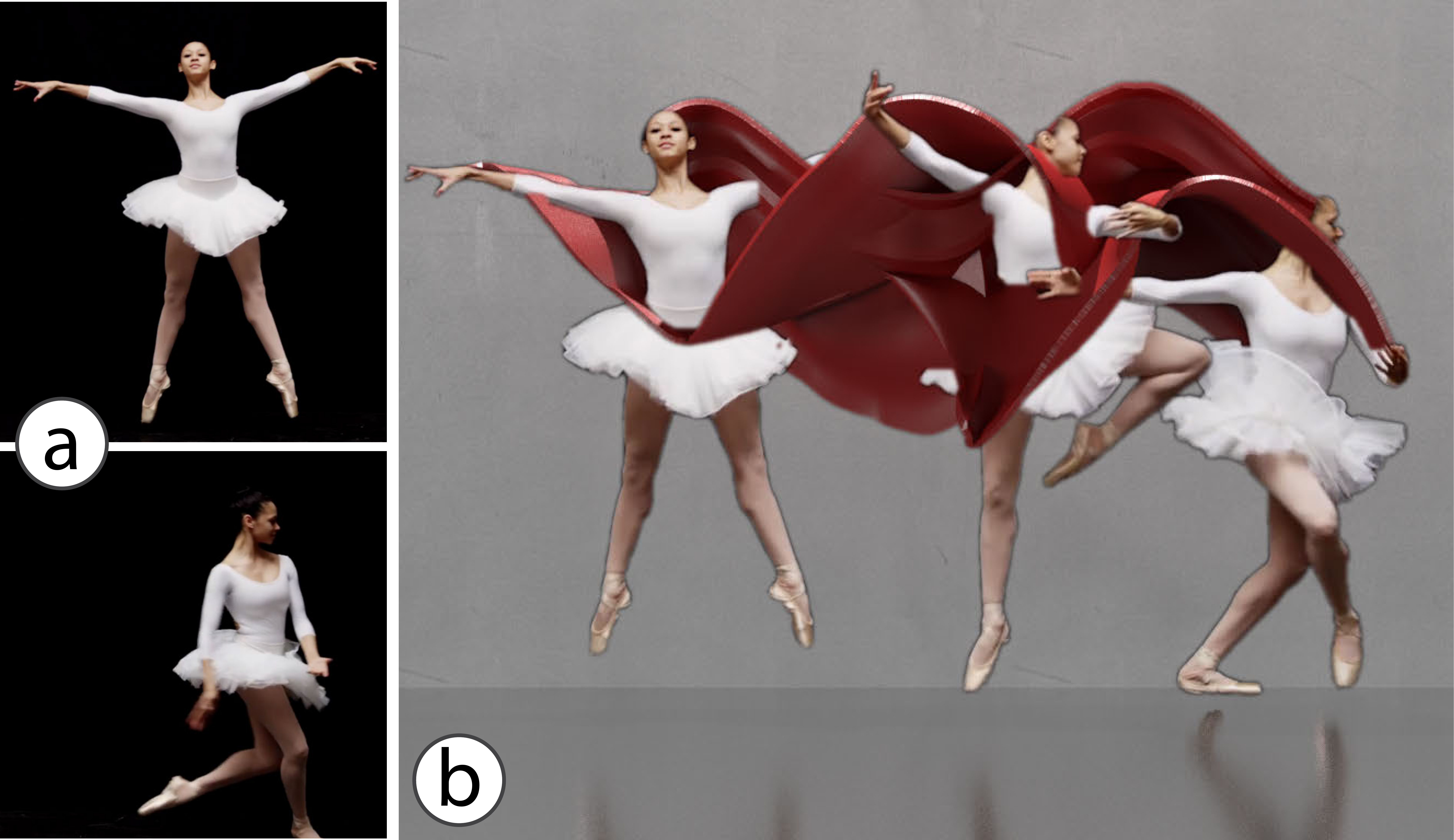}
 	%\vspace{-15pt}
	\caption{The \textit{Ballet-2} sculpture (material: leather; rendered body parts: body and arms). (a) First and final frames. (b) The \summaryname rendered in a synthetic scene.}
     \vspace{-5pt}
	\label{fig:dancing}
 \end{figure}

In \emph{Tennis} (\fig{fig:comparison} bottom), the sculpture highlights bending of the arm during the serve, which is not easily visible from 2D or 2.5D visualizations (also shown in \fig{fig:comparison} bottom). Similarly, in \emph{Ballet-2}~\cite{ballet} (\fig{fig:dancing}), a sinusoidal 3D surface emerges from the motion of the ballerina's right arm, again absent in the 2D or 2.5D visualizations.

\begin{figure*}[t]
    \centering
    \includegraphics[width=\textwidth]{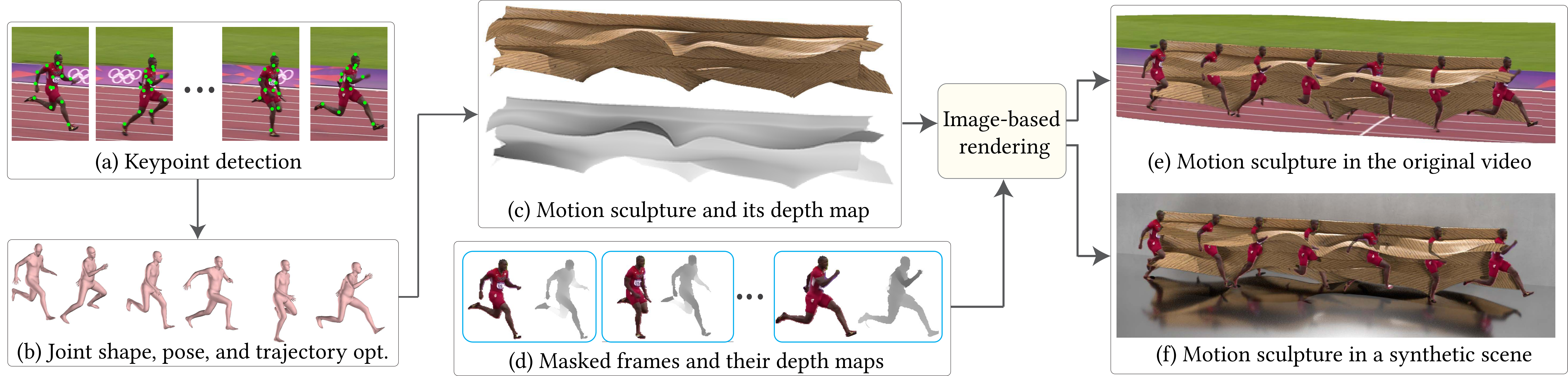}
    %\vspace{-10pt}
    \caption{\modelname workflow. Given an input video, we first detect 2D keypoints for each video frame (a), and then estimate a 3D body model that represents the person's overall shape and its 3D poses throughout the video, in a temporally coherent manner (b). The \summaryname is formed by extracting 3D skeletons from the estimated, posed shapes and connecting them (c). Finally, by jointly considering depth of the sculpture (c) and the human bodies (d), we render the sculpture in different styles, either into the original video (e) or a synthetic scene (f).}
     \label{fig:pipeline}
    %  \vspace{-10pt}
\end{figure*}

\section{Algorithm for Generating Motion Sculptures}

The algorithm behind \modelname consists of several steps illustrated in \fig{fig:pipeline}. In short, our algorithm (a) first detects the human body and its 2D pose (represented by a set of keypoints) in each frame, (b) recovers a 3D body model that represents the person's overall shape and its 3D poses across the frames, in a temporally coherent manner, (c) extracts a 3D skeleton from the 3D model and sweeps it through the 3D space to create an initial \summaryname, and finally, (d-f) renders the sculpture in different styles, together with the human, while preserving the depth ordering.   

\subsection{2D Keypoint Detection}

The 2D body pose in each frame, represented by a set of 2D keypoints, is estimated using OpenPose~\cite{cao2017realtime}. Each keypoint is associated with a joint label (\eg, left wrist, right elbow) and its 2D position in the frame.

%AO: maybe only mention OpenPose once or twice -- otherwise it'll seem too derivative of it. E.g. instead of calling it OpenPose keypoint detections, just talk about the keypoints. You can also use "\cite{cao2017realtime}" at later points in the text, or say we use a keypoint-based pose reconstruction method
While keypoints detected in a single image are typically accurate, inherent ambiguity in the motion of a human body sometimes leads to temporal inconsistency, \eg, the left and right shoulders flipping between adjacent frames. We address this problem by imposing temporal coherency between detections in adjacent frames. Specifically, we use a Hidden Markov Model (HMM), where the per-frame detection results are the observations. We compute the maximum marginal likelihood estimate of each joint's location at a specific timestamp, while imposing temporal smoothness (see the supplementary material for more details). 
%For frames, for which OpenPose could not detect the joints, we linearly interpolate their locations from adjacent frames before running the HMM.

We develop a simple interface (\fig{fig:userinterface}a), where the user can browse through the detection results (overlaid on the video frames) and indicate whether the detected joints are all correct in a given frame. The frames labeled correct are then used as constraints in another HMM inference procedure. Three or four labels are usually sufficient to correct all the errors in a video of 100 frames.

\subsection{From 2D Keypoints to 3D Body Over Time}
\label{sec:3dpose}

Given the detected 2D keypoints, our goal now is to fit a 3D model of the body in each frame. We want temporally consistent configurations of the 3D body model that best match its 2D poses (given by keypoints). That is, we opt to minimize the re-projection error, \ie, the distance between each 2D keypoint and the 3D-to-2D projection of the mesh vertices that correspond to the same body part. 

We use the SMPL~\cite{loper2015smpl} body model that consists of a canonical mesh and a set of parameters that control the body shape, pose, and position. Specifically, the moving body is represented by shape parameters $\beta$, per-frame pose $\theta^t$, and global translation ${T^t}$. We estimate these parameters for each of the $N$ frames by minimizing the following objective function:
\begin{align}
\label{eqn:core}
%\mathcal{L}\left(\{T^t\}, \{\theta^t\}, \beta\right)
\mathcal{L}&\left(\{T^t\}, \{\theta^t\}, \beta\right) = \sum_{t=1}^N\left(\mathcal{L}_\text{data}\left(T^t, \theta^t, \beta\right) + \alpha_1\mathcal{L}_\text{prior}\left(\theta^t, \beta\right)\right)\nonumber\\
&+ \alpha_2\sum_{t=1}^{N-1}\mathcal{L}_\text{temporal}\left(T^t, T^{t+1}, \theta^t, \theta^{t+1}, \beta\right).
\end{align}

The data term $\mathcal{L}_\text{data}$ encourages the projected 3D keypoints in each frame to be close to the detected 2D keypoints. $\mathcal{L}_\text{prior}$ is a per-frame prior defined in~\cite{bogo2016keep}, which imposes priors on the human pose as well as joint bending, and additionally penalizes mesh interpenetration. Finally, $\mathcal{L}_\text{temporal}$ encourages the reconstruction to be smooth by penalizing change in the human's global translations and local vertex locations. $\alpha_i$ are hand-chosen constant weights that maintain the relative balance between the terms. This formulation can be seen as an extension of SMPLify~\cite{bogo2016keep}, a single-image 3D human pose and shape estimation algorithm, to videos. The optimization is solved using~\cite{loper2014opendr}. See the supplementary material for the exact term definitions and implementation details.

\subsection{Generating the Sculpture}

\begin{figure}[t]
    \centering
    \includegraphics[width=1\linewidth]{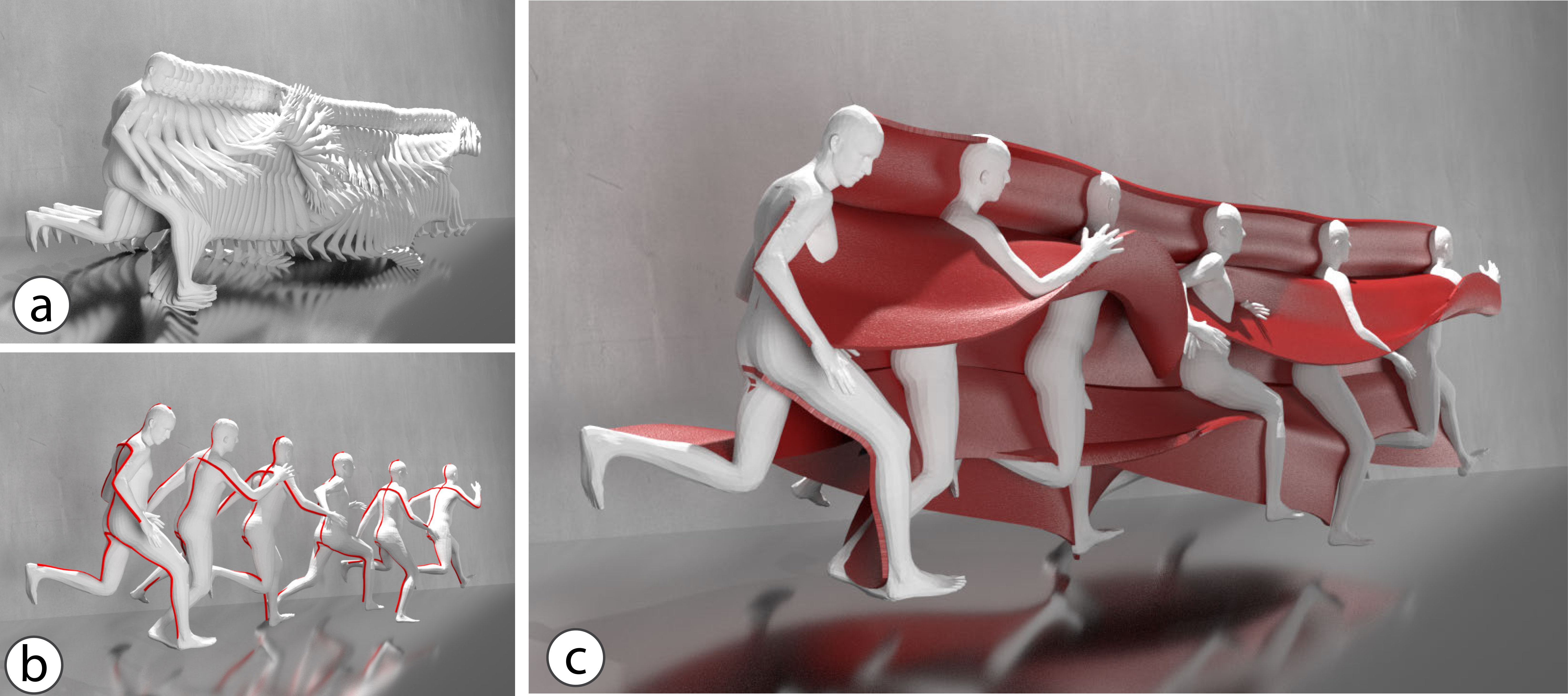}
    %\vspace{-12pt}
    \caption{
    Sculpture formation. (a) A collection of shapes estimated from the \textit{Olympic} sequence (see \fig{fig:teaser}). (b) Extracted 3D surface skeletons (marked in red). (c) An initial \summaryname is generated by connecting the surface skeletons across all frames.}
     \label{fig:sculp-formation}
     %\vspace{-1pt}
\end{figure}

With a collection of 3D body shapes (\fig{fig:sculp-formation}a), we create a space-time sweep by extracting the reconstructed person's skeleton from the 3D model in each frame (marked red on the shapes in \fig{fig:sculp-formation}b) and connecting these skeletons across all frames (\fig{fig:sculp-formation}c). This space-time sweep forms our initial \summaryname. 

 \begin{figure*}[t!]
 	\includegraphics[width=\textwidth]{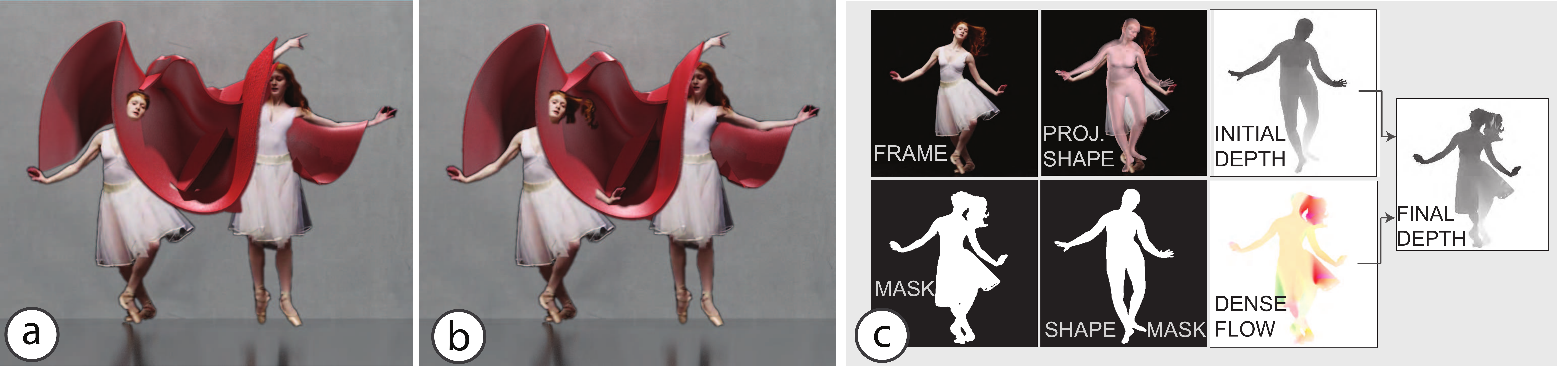}
 	%\vspace{-5pt}
	\caption{Flow-based refinement. (a) \Summaryname rendering without refinement; gaps between the human and the sculpture are noticeable. (b) Such artifacts are eliminated with our flow-based refinement scheme. (c) We first compute a dense flow field between the frame silhouette (bottom left) and the projected 3D silhouette (bottom middle). We then use this flow field (bottom right) to warp the initial depth (top right; rendered from the 3D model) and the 3D sculpture to align them with the image contents.} %and propagate  depth values to not modeled by the initial 3D shape.}
 	%\vspace{-5pt}
	\label{fig:alignment}
 \end{figure*}

% AO: can we cut this paragraph? Not sure I completely get it. 
% TD: take a look now.
%Note that our 3D skeletons are \emph{surface skeletons}, \ie, consist of a set of contours over the surface of the shape, rather than simply connecting lines between the joints. The surface skeleton better captures the body's shape as they better align with the body's occlusion boundary. In practice, we extract the surface skeletons by projecting the standard 3D skeleton onto the back surface of the shape (\fig{fig:sculp-formation}b). 
% AO: I cut the reference to SMPL

% AO: maybe we can retitle this something more general, like "Rendering motion sculptures"?
% AO: as a general comment I think we can cut a lot of details here.
\section{Refining and Rendering Motion Sculptures}
\label{sec:embed-sculp}
In order to achieve artifact-free and vivid renderings, we still have several remaining issues to resolve. First, a generic 3D body model (such as the one that we use) cannot accurately capture an individual's actual body shape In other words, it lacks important structural details, such as fine facial structure, hair, and clothes. Second, our reconstruction only estimates the geometry, but not the texture. Texture mapping from 2D to 3D under occlusion itself is a challenging task, even more so when the 3D model does not cover certain parts of the body. \fig{fig:two-worlds}a illustrates these challenges: full 3D rendering lacks structural details and results in noticeable artifacts.

Our approach is inserting the 3D \summaryname back into the original 2D video, rather than mapping the 2D contents from the video to the 3D scene. This allows us to preserve the richness of information readily available in the input video (\fig{fig:two-worlds}c) without modeling fine-scale (and possibly idiosyncratic) aspects of the 3D shape.

\subsection{Depth-Aware Composite of 3D Sculpture and 2D Video}

\begin{figure}[t]
 	\includegraphics[width=\columnwidth]{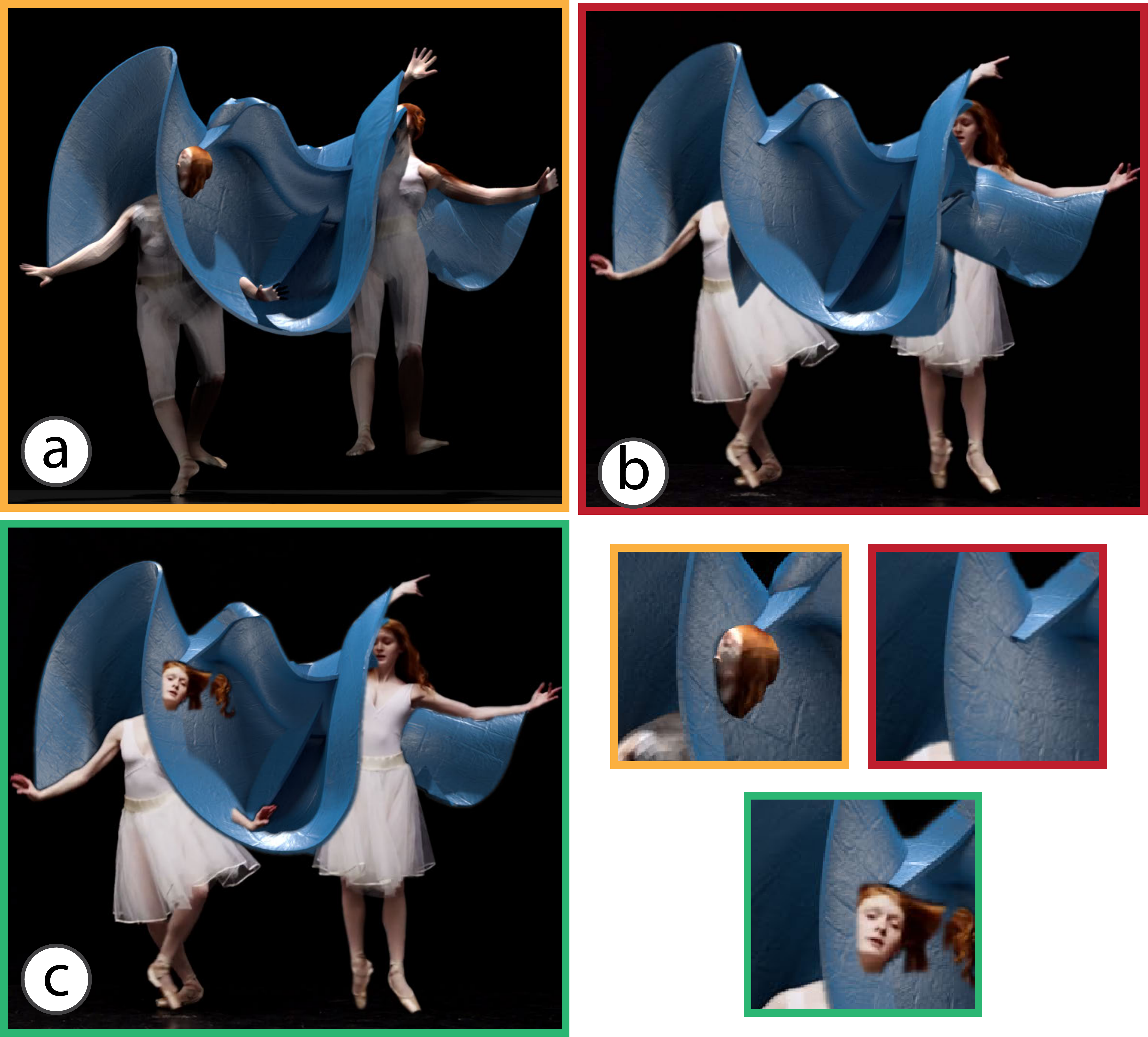}
 	%\vspace{-8pt} 
 	\caption{(a) Full 3D rendering using textured 3D human meshes exposes artifacts and loses important appearance information, \eg, the ballerina's hair and dress. (b) Simply placing the sculpture on top of the frame discards the information about depth ordering. (c) Our 3D-aware image-based rendering approach preserves the original texture as well as appearance, and reveals accurate 3D occlusion relationship.}
 	% fix colors later
	\label{fig:two-worlds}
	%\vspace{-10pt}
 \end{figure}

As can be seen in \fig{fig:two-worlds}b, naively superimposing the rendered 3D sculpture onto the video results in a cluttered visualization that completely disregards the 3D spatial relationships between the sculpture and the object. Here, the person's head is completely covered by the sculpture, making shape and motion very hard to interpret.  %uses optical flow to refine the sculpture and object depth (\fig{fig:pipeline}d) and 
We address this issue and produce depth-preserving composites such as the one in \fig{fig:two-worlds}c.

%infer the first analyze the 2D video frame and extract a
To accomplish this, we estimate a depth map of the person in each video frame. For each frame and each pixel, we then determine if the person is closer to or farther away from the camera than the sculpture by comparing the sculpture's and person's depth values at that pixel (the sculpture depth map is automatically given by its 3D model). 
%If the person is closer to the camera, the person is rendered in front, otherwise we render the motion sculpture in front. 
We then render at each pixel what is closer to the camera, giving us the result shown in \fig{fig:two-worlds}c.

%the girl's arm is thinner than the one in the 3D model,
%and there is a gap.  
% \input{figText/alignment}

\subsection{Refinement of Depth and Sculpture}
While the estimated sculpture is automatically associated with a depth map, this depth map rarely aligns perfectly with the human silhouette. Furthermore, we still need to infer the human's depth map in each frame for depth ordering. As can be seen in \fig{fig:alignment}c, the estimated 3D body model  provides only a rough and partial estimation of the human's depth due to misalignment and missing 3D contents (\eg, the skirt or hair).  A rendering produced with these initial depth maps leads to visual artifacts, such as wrong depth ordering and gaps between the sculpture and the human (\fig{fig:alignment}a),

To eliminate such artifacts, we extract foreground masks of the human across all frames (using Mask R-CNN~\cite{he2017mask} followed by $k$-NN matting~\cite{Chen:2012:KM}), and refine the human's initial depth maps as well as the sculpture as follows.
%\begin{enumerate}
% AO: I think this is too much detail. can we cut some of this down?
% TD: I think people in UIST are not all computer vision people that is the reason for the details...

For refining the object's depth, we compute dense matching, \ie, optical flow~\cite{liu:opticalflow}, between the 2D foreground mask and the projected 3D silhouette. We then propagate the initial depth values (provided by the estimated 3D body model) to the foreground mask via warping with optical flow. If a pixel has no depth after warping, we copy the depth of its nearest neighbor pixel that has depth. This approach allows us to approximate a complete depth map of the human. As shown in \fig{fig:alignment}c, the refined depth map has values for the ballerina's hair and skirt, allowing them to emerge from the sculpture (compared with the hair in \fig{fig:alignment}a).
    
    % AO: can you clarify what "editing" means here? That initially gave me the impression that the user could edit it. I think we can cut a lot of this (I was confused about the part about jittering, too).
    % I think that some of the reviewers had problems with this, too, since they commented that there were a lot of different masking methods (e.g. Mask R-CNN and many others). Streamlining the presentation by cutting stuff could help.
    % TD: take a look now
For refining the sculpture, recall that a \summaryname is formed by a collection of surface skeletons. We use the same flow field as above to warp the image coordinates of the surface skeleton in each frame. Now that we have determined the skeletons' new 2D locations, we edit the \summaryname in 3D accordingly\footnote{We back-project the 2D-warped surface skeletons to 3D, assuming the same depth as before editing. Essentially, we are modifying the 3D sculpture in only the $x$- and $y$-axes. To compensate for some minor jittering introduced, we then smooth each dimension with a Gaussian kernel.}. After this step, boundary of the sculpture, when projected to 2D, aligns well with the 2D human mask.

% Thus, we repeat the steps from 'extracting and connecting skeletons to create sweep', i.e. we back-project the warped surface contour to 3D and sweep it across all frames. 

% We are editing the 3D sculpture in the $x$- and $y$-axes such that its boundary, when projected to 2D, aligns well with the source 2D images. To do this, w
%\end{enumerate}

% Essentially, we are editing the 3D sculpture in the $x$- and $y$-axes such that its boundary, when projected to 2D, aligns well with the original 2D images. To do this, we compute the dense optical flow~\cite{liu:opticalflow} between a foreground mask of the human and the projected, rendered 3D silhouette and then use it to warp the image coordinates of the 3D surface contours (that form the sculpture).  

% AO: can we just say we get foreground masks using Mask R-CNN (no need to say "efficiently and accurately". This is a little confusing too because we already talked about flow and depth. Losing track of how we compute masks, depth, and flow. I think shortening this section would help address this. 
%\ignore{We run Mask R-CNN~\cite{he2017mask} on each frame to compute loose foreground masks, which we then erode to produce the corresponding overtight masks. Combining the loose and overtight masks produces trimaps, which are then fed to kNN matting~\cite{Chen:2012:KM}.}
% AO: what are trimaps?
% TD: I just removed all this part all together...

\section{User Studies}
\label{sec:userstudy}
%\tali{Missing general discussion and details.}
We conducted several user studies to \change{compare how well motion and shape are perceived from different visualizations, and evaluate the stylistic settings provided by our interface.}

%We conducted user studies to {\color{red}(i) find out which visualization \emph{overall} conveys the clearest information about motion,} (ii) compare how easily users can perceive \emph{particular} information about shape and motion from different visualizations, and (iii) evaluate the the stylistic settings provided by our interface.

\subsection{\SummaryName \emph{vs.\xspace} Stroboscopic \emph{vs.\xspace} Shape-Time}

\begin{table}[t]
    \centering
    \setlength{\tabcolsep}{3pt}
    \begin{tabular}{lcccccc}
    \toprule
    & Bal1 & Bal2 & Jog & Olym & Walk & {\bf Avg} \\
    \midrule
    %Prefer Ours to Strobo. & 70 & 52 & 60 & 70 & 57 & {\bf 62} \\
    %Prefer Ours to \cite{freeman:shapetime} & 58 & 68 & 52 & 55 & 57 & {\bf 58} \\
    Prefer Ours to Strobo & 92 & 75 & 69 & 69 & 58 & {\bf 73} \\
    Prefer Ours to \cite{freeman:shapetime} & 81 & 78 & 78 & 83 & 61 & {\bf 76} \\
    \bottomrule
    \end{tabular}
    %\vspace{-5pt}
    \vspace{2pt}
    \caption{Percentage. We conducted human studies to compare our visualization with stroboscopic and shape-time photography~\protect\cite{freeman:shapetime}. Majority of the subjects suggested that ours conveys more motion  information.}
    \label{tbl:comparison}
    \vspace{-5pt}
\end{table}
%\tali{describe in more details the protocol}

We asked the participants to rate how well motion information is conveyed in \summarynames, stroboscopic photography, and shape-time photography~\cite{freeman:shapetime} for five clips. An example is shown in  \fig{fig:comparison}, and the full set of images used in our user studies is included in the supplementary material. 

 \change{In the first test, we presented the raters with two different visualizations (ours \vs a baseline), and asked ``which visualization provides the clearest information about motion?''. We collected responses from 51 participants with no conflicting interests for each pair of comparison. 77\% of the responses preferred our method to shape-time photography, and 67\% preferred ours to stroboscopic photography.

 In the second study, we compared how easily users can perceive \emph{particular} information about shape and motion from different visualizations. To do so, we asked the following clip-dependent} questions: ``which visualization helps more in seeing:
\begin{itemize}
    \myitem the arm moving in front the body (\textit{Ballet-1}),
    \myitem the wavy and intersecting arm movement (\textit{Ballet-2}),
    \myitem the wavy arm movement (\textit{Jogging} and \textit{Olympics}), or
    \myitem the person walking in a U-shape (\textit{U-Walking}).''
\end{itemize}

\change{We collected 36 responses for each sequence. As shown in \tbl{tbl:comparison}, on average, the users preferred our visualization over the alternatives 75\% of the time.  The questions above are intended to focus on the salient 3D characteristics of motion in each clip, and the results support that our visualization conveys them better than the alternatives. For example, in \textit{Ballet-1} (\fig{fig:comparison} top), our \summaryname visualizes the out-of-plane sinusoidal curve swept out by the ballerina's arm, whereas both shape-time and stroboscopic photography show only the in-plane motion. Furthermore, our \summaryname shows the interactions between the left and right arms.}
%\mytodo{I would perfer avoiding long subjective explanation. I would cut the next sentence}
%Thus, the viewer can read out that the ballerina's left arm penetrates the space once traversed by her right arm, whereas this is not obvious in both the shape-time and stroboscopic photography.
% Xiuming: I think this observation is pretty objective? But I'm fine with cutting it out, as it's too long

%Thus, our method is better suited for revealing such spatial-temporal interactions over time. 

% Details <shorten by half at least>: The input format for all three methods is different, we proceeded as following when preparing the source images:  Because shape-time photography works on RGB-depth image pairs, we fed our approximated depth maps to the algorithm in addition to the original video. Directly applying the method in \cite{freeman:shapetime} led to a considerable number of artifacts, especially near depth boundaries, perhaps due to the complexity of the scenes. We therefore adapted the model of Freeman and Zhang~\cite{freeman:shapetime} to normal videos (rather than low frame-rate videos or hand-selected frames, as in~\cite{freeman:shapetime}) by augmenting it with the texture smoothness prior of~\cite{pritch:shiftmap} and Potts smoothness terms. This removes artifacts resulting from depth errors and adaptively selects a sparser set of output frames, making results significantly easier to interpret. %We found that 

\begin{figure}[t]
    \centering
    %\begin{subfigure}{\linewidth}
    \includegraphics[width=\linewidth]{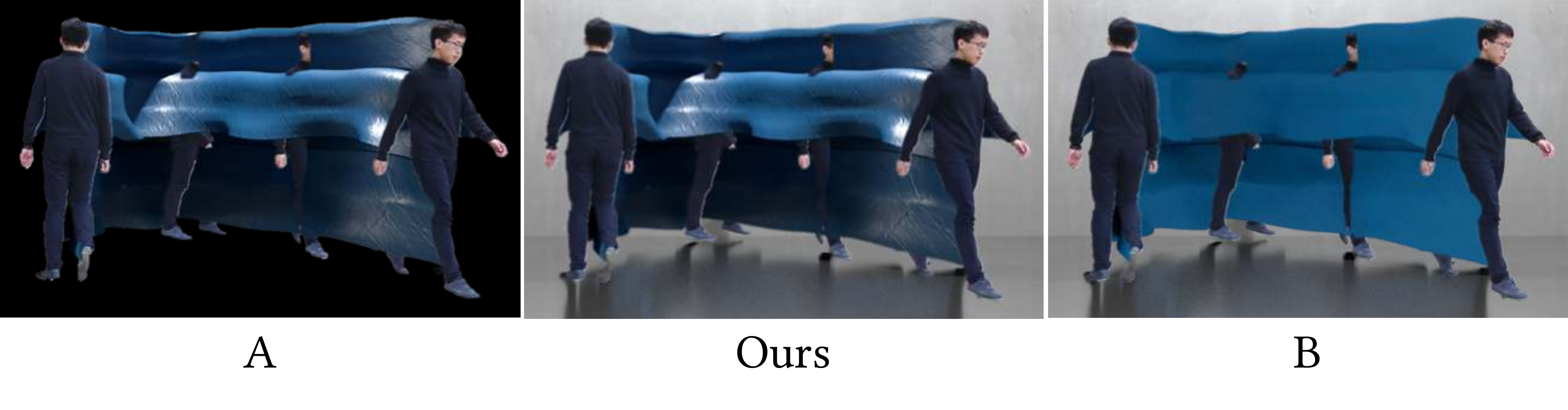}
    %\vspace{-20pt}
    % \caption{%\tianfan{The difference between ``ours'' and B is minor. It is better to show a zoom-up.} 
    % Our rendering (middle) with two variants: (A) without reflections and shadow, and (B) without localized lighting.} % The first study investigates whether reflections and shadows enhance the renderings by asking the users to pick from A and ours, and the second study, asking the users to pick from B and ours, examines whether localized lighting produces stronger 3D cues than ambient lighting does.}
    %\label{fig:human-studies-fig}
    %\end{subfigure}
    
    %\begin{subfigure}{\linewidth}
    %\centering
    %\vspace{10pt}
    \setlength{\tabcolsep}{1pt}
    \begin{tabular}{lccccccccc}
    \toprule
     & Tenn & Bal1 & Bal2 & Jump & Walk & Olym & Dunk & {\bf Avg} \\
    \midrule
    Prefer Ours to A & 93 & 63 & 86 & 83 & 83 & 93 & 73 & {\bf 82} \\
    Prefer Ours to B & 78 & 94 & 84 & 78 & 91 & 78 & 79 & {\bf 84} \\
    \bottomrule
    \end{tabular}
    %\vspace{-5pt}
    %\caption{}%The human studies results demonstrate that most raters agree with our design choices: on average, 82\% the use of reflections and shadow, and 84\% prefer our localized lighting.}
    %\label{tbl:human-studies-tbl}
    %\end{subfigure}
    \vspace{2pt}
    \caption{We conducted human studies to justify our artistic design choices. Top: sample stimuli used in the studies -- our rendering (middle) with two variants, without reflections (A) and without localized lighting (B). Bottom: percentage; most of the subjects agreed with our choices.}
    \label{fig:human-studies}
    %\vspace{-8pt}
\end{figure}
\begin{figure}[t]
    \centering
    \includegraphics[width=\linewidth]{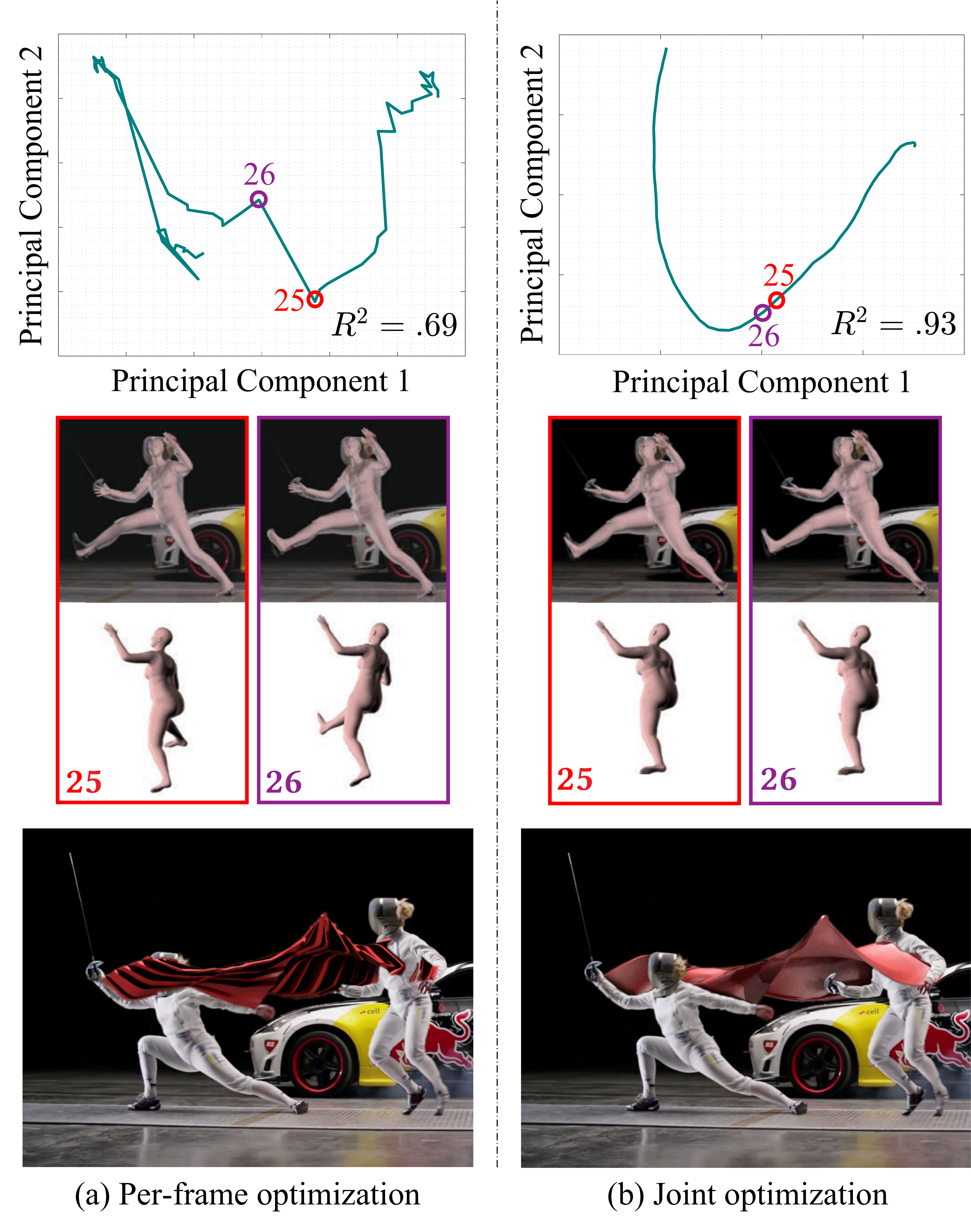}
    %\vspace{-20pt}
    \caption{Per-frame \vs joint optimization. (a) Per-frame optimization produces drastically different poses between neighboring frames (\eg, from frame 25 [red] to frame 26 [purple]). The first two principal components explain only 69\% of the pose variance. (b) On the contrary, our joint optimization produces temporally smooth poses across the frames. The same PCA analysis reveals that the pose change is gradual, lying on a 2D manifold with 93\% of the variance explained.}
     \label{fig:pose}
\end{figure}

\subsection{Effects of Lighting and Floor Reflections}
To avoid exposing too many options to the user, we conducted a user study to decide (i) whether floor reflections are needed in our synthetic-background rendering, and (ii) whether localized or global lighting should be used.
%In this study, we evaluate different stylistic settings provided by our UI, such as the use of (i) reflections and shadows, and (ii) localized lighting vs. global lighting.
The raters were asked which rendering is more visually appealing: with \vs without reflections (Ours \vs A), and using localized \vs ambient lighting (Ours \vs B).

\fig{fig:human-studies} shows the results collected from 20-35 responses for each sequence on Amazon Mechanical Turk, after filtering out workers who failed our consistency check. Most of the raters preferred our rendering with reflections plus shadows (82\%) and localized lighting (84\%) to the other options. We thus use these as the standard settings in our user interface. % but still allow the user to customize them on demand.

%\tali{complete \fig{fig:ablation-smoothness} and discuss results}
% We compare our full model with a few variants, both quantitatively and qualitatively, to demonstrate the advantages of our technical innovations: our model (1) without temporal smoothness, (2) without editing the sculpture, or (3) without editing the depth maps.
\section{Technical Evaluation}
\label{sec:result}
%\tali{Missing general discussion and details.}

%Printing time: total 9hrs (2 sculpture printed at the same time) 
%printer: Stratasys obj260
%Material : Rigur (RGD450)

%\vspace{-5pt}
\label{sec:eval-pipeline-comp}
We conducted experiments to evaluate our two key technical components: (i) 3D body estimation over time, and (ii) flow-based refinement of depth and sculpture. 

\subsection{Estimating Geometry Over Time}

In our first evaluation, we compared our approach that estimates the correct poses by considering change across multiple frames against the pose estimation of SMPLify~\cite{bogo2016keep}, in which the 3D body model is estimated in each frame independently. \fig{fig:pose}a shows the output of SMPLify, and \fig{fig:pose}b shows our results. The errors in the per-frame estimates and the lack of temporal consistency in \fig{fig:pose}a resulted in a jittery, disjoint sculpture. In contrast, our approach solved for a single set of shape parameters and smoothly varying pose parameters for the entire sequence, and hence produced significantly better results.

To quantitatively demonstrate the effects of our approach on the estimated poses, we applied Principal Component Analysis (PCA) to the 72D pose vectors, and visualized the pose evolution in 2D in~\fig{fig:pose}. In SMPLify (\fig{fig:pose}a), there is a significant discrepancy between poses in frames 25 and 26: the human body abruptly swings to the right side. In contrast, with our approach, we obtained a smooth evolution of poses (\fig{fig:pose}b).

\subsection{Flow-Based Refinement}

As discussed earlier, because the 3D shape and pose are encoded using low-dimensional basis vectors, perfect alignment between the projected shape and the 2D image is unattainable. These misalignments show up as visible gaps in the final renderings. However, our flow-based refinement scheme can significantly reduce such artifacts (\fig{fig:alignment}b). 

To quantify the contribution of the refinement step, we computed Intersection-over-Union (IoU) between the 2D human silhouette and projected silhouette of the estimated 3D body. \tbl{tbl:iou} shows the average IoU for all our sequences, before and after flow refinement. As expected, the refinement step significantly improves the 3D-2D alignment, increasing the average IoU from 0.61 to 0.94. After hole filling with the nearest neighbor, the average IoU further increases to 0.96.

% \begin{table}
%  	\centering
%     \begin{tabular}{lccccccccc}
%     \toprule
%      & Tenn & Fenc & Bal1 & Bal2 & Jump & Walk & Olym & Avg \\
%     \midrule
%     Raw & 0.56 & 0.87 & 0.54 & 0.60 & 0.57 & 0.68 & 0.65 & 0.61 \\
%     Warp & 0.97 & 0.93 & 0.93 & 0.93 & 0.98 & 0.95 & 0.86 & 0.94 \\
%     Warp \\+ HF & {\bf 0.98} & {\bf 0.99} & {\bf 0.96} & {\bf 0.96} & {\bf 0.99} & {\bf 0.96} & {\bf 0.92} & {\bf 0.96} \\
%     \bottomrule
%     \end{tabular}
%     \vspace{10pt}
%     \caption{IoU between the binarized human depth maps and human silhouettes before warping, after warping, and after  additional hole filling (HF). Flow-based refinement of the approximated human depth maps leads to better alignment with the original images, which in turn improves the final renderings.}
%     \label{tbl:iou}
% \end{table}

\begin{table}[t]
 	\centering
 	\setlength{\tabcolsep}{2.5pt}
    \begin{tabular}{cccccccccc}
    \toprule
     & Tenn & Fenc & Bal1 & Bal2 & Jump & Walk & Olym & Avg \\
    \midrule
    Raw & .56 & .87 & .54 & .60 & .57 & .68 & .65 & .64 \\
    Warp & .97 & .93 & .93 & .93 & .98 & .95 & .86 & .94 \\
    Warp+HF & {\bf .98} & {\bf .99} & {\bf .96} & {\bf .96} & {\bf .99} & {\bf .96} & {\bf .92} & {\bf .97} \\
    \bottomrule
    \end{tabular}
    \vspace{2pt}
    \caption{IoU between human silhouettes and binarized human depth maps before warping, after warping, and after additional hole filling with nearest neighbor (HF). Flow-based refinement leads to better alignment with the original images and hence improves the final renderings.}
    \label{tbl:iou}
    %\vspace{-10pt}
\end{table}

% \begin{table}
%  	\centering
%     \begin{tabular}{lccc}
%     \toprule
%      & Raw & Warped & Warped + Hole Filling\\
%     \midrule
%     Federer    &  0.56  & 0.97 & 0.98\\
%     Fencing  &  0.87   & 0.93 & 0.99 \\
%     Ballet-1    &  0.60  & 0.93 & 0.96\\
%     Ballet-2   &  0.54   & 0.93 & 0.96\\
%     Jumping    &  0.57  & 0.98 & 0.99\\
%     U-Walking  &  0.68   & 0.95 & 0.96\\
%     Jogging    &  0.66  & 0.97 & 0.98\\
%     Olympic  &  0.65   & 0.86 & 0.92\\
%     \hline
%     Average &  0.61   & 0.94& 0.96\\
%     \bottomrule
%     \end{tabular}
%     \vspace{10pt}
%     \caption{IoU between the binarized human depth maps and human silhouettes before warping, after warping, and after  additional hole filling. Flow-based refinement of the approximated human depth maps leads to better alignment with the original images, which in turn improves the final renderings.}
%     \label{tbl:iou}
% \end{table}
\section{Implementation Details}
\label{sec:impl_details}

We rendered our scenes using Cycles in Blender. \change{It took a Stratasys J750 printer around 10 hours to 3D print the sculpture shown in \fig{fig:teaser}d ($\sim$30cm long).} To render realistic floor reflections in synthetic scenes, we coarsely textured the 3D human with simple ray casting\change{: we cast a ray from each vertex on the human mesh to the estimated camera, and colored that vertex with the RGB value of the intersected pixel. Intuitively, this approach mirrors texture of the visible parts to obtain texture for the occluded parts. The original texture for sculptures (such as the sculpture texture in \fig{fig:walking}) was computed similarly, except that when the ray intersection fell outside the (eroded) human mask, we took the color of the intersection's nearest neighbor inside the mask to avoid colors being taken from the background. As an optional post-processing step, we smoothed the vertex colors over each vertex's neighbors.} Other sculpture texture maps (such as wood) were downloaded from \url{poliigon.com}.

To render a \summaryname together with the human figures, we first rendered the 3D sculpture's RGB and depth images as well as the human's depth maps using the recovered camera. We then composited together all the RGB images by selecting, for each pixel, the value that is the closest to the camera, as mentioned before. Due to the noisy nature of the human's depth maps, we used a simple Markov Random Field (MRF) with Potts potentials to enforce smoothness during this composition.

%Our hybrid setup thus requires only little to zero human texturing (for reflections) to produce high-quality 3D rendering.

For comparisons with shape-time photography~\cite{freeman:shapetime}, because it requires RGB and depth image pairs as input, we fed our refined depth maps to the algorithm in addition to the original video. Furthermore, shape-time photography was not originally designed to work on high-frame-rate videos; directly applying it to such videos leads to a considerable number of artifacts. We therefore adapted the algorithm to normal videos by augmenting it with the texture smoothness prior in~\cite{pritch:shiftmap} and Potts smoothness terms. 
%This significantly improved their results and removes artifacts resulting from depth errors and adaptively selects a sparser set of output frames, making results significantly easier to interpret. %We found that 
\begin{figure}[t]
    \centering
    \includegraphics[width=\linewidth]{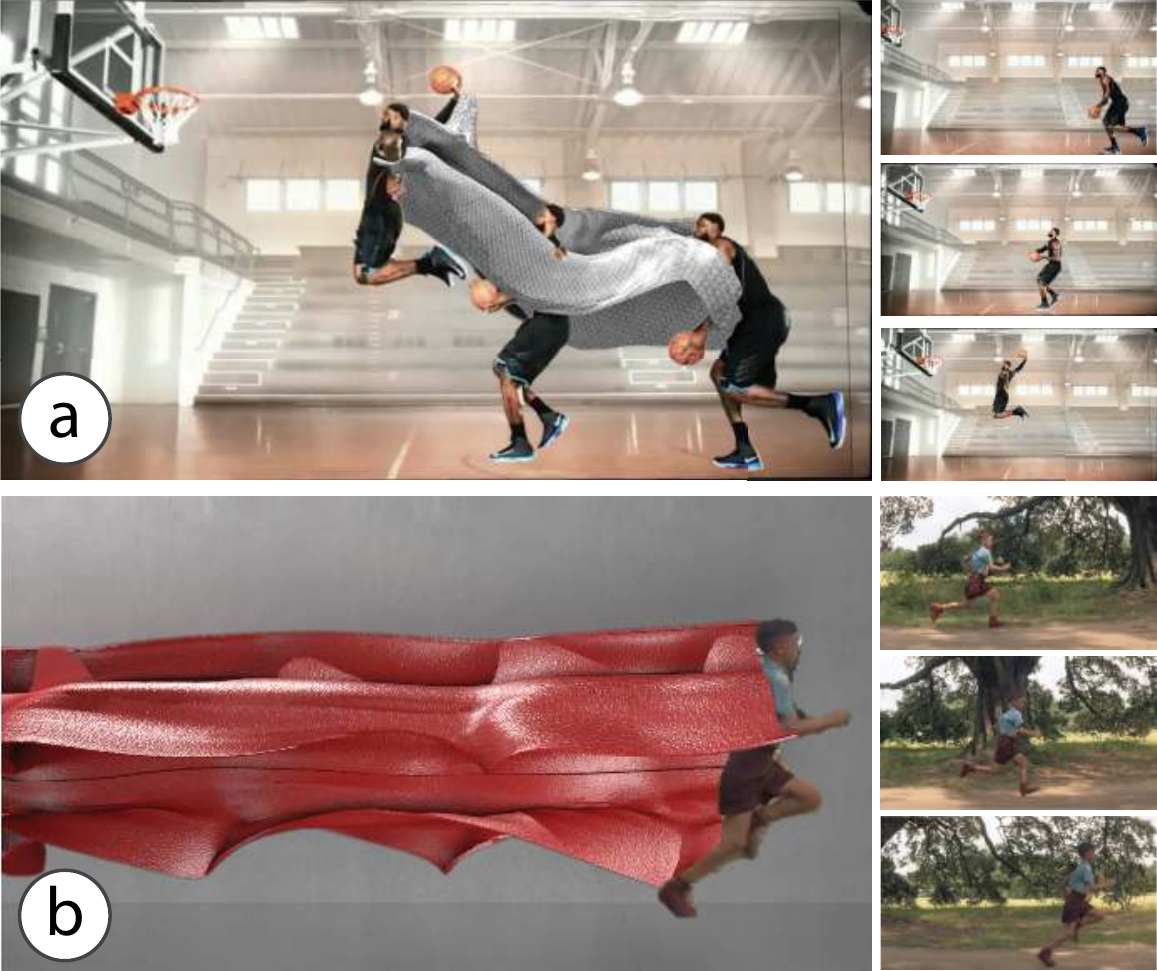}
    % \missingfigure{Figure for horses.}
    % \vspace{-20pt}
    \caption{
    %\tianfan{Don't like the name ``Run, Forect, Run!''. Prefer ``Running Forest''}
    %\xiuming{This scene is actually quite widely known as ``Run, Forrest, Run!'': https://www.google.com/search?q=run+forrest+run&rlz=1C5CHFA_enUS723US723&source=lnms&tbm=isch&sa=X&ved=0ahUKEwipiuuqv-rYAhXtk-AKHYeJDcMQ_AUICigB&biw=2560&bih=1272.}
    \Summarynames for videos captured by moving cameras.}
     \label{fig:movcam}
    %  \vspace{-20pt}
\end{figure}

\section{Extensions}

We extend our model to handle camera motion and generate non-human \summarynames.

\subsection{Handling Camera Motion}

As an additional feature, we extend our algorithm to also handle camera motion. One approach for doing so is to stabilize the background in a pre-processing step, \eg, by registering each frame to the panoramic background~\cite{brown2007automatic}, and then applying our system to the stabilized video. This works well when the background is mostly planar. Example results obtained with this approach are shown for the \emph{Olympic} and \emph{Dunking} videos, in \fig{fig:teaser} and \fig{fig:movcam}a, respectively.

However, for more complex scenes containing large variations in depth, this approach may result in artifacts due to motion parallax. Thus, for general cases, we use an off-the-shelf Structure-from-Motion (SfM) software~\cite{schoenberger2016sfm} to estimate the camera position at each frame and then compensate for it. More specifically, we estimate the human's position relative to the moving camera, and then offset that position by the camera position given by SfM. An example of this approach is \emph{Run, Forrest, Run!}, shown in \fig{fig:movcam}b. As can be seen, our method  works well on this challenging video, producing a \summaryname spanning a long distance (\fig{fig:movcam}b has been truncated due to space limit, so the actual sculpture is even longer; see the supplementary video).

\subsection{Non-Human \SummaryNames}
\label{sec:numhuman}

\begin{figure}[t]
    \centering
    \includegraphics[width=\linewidth]{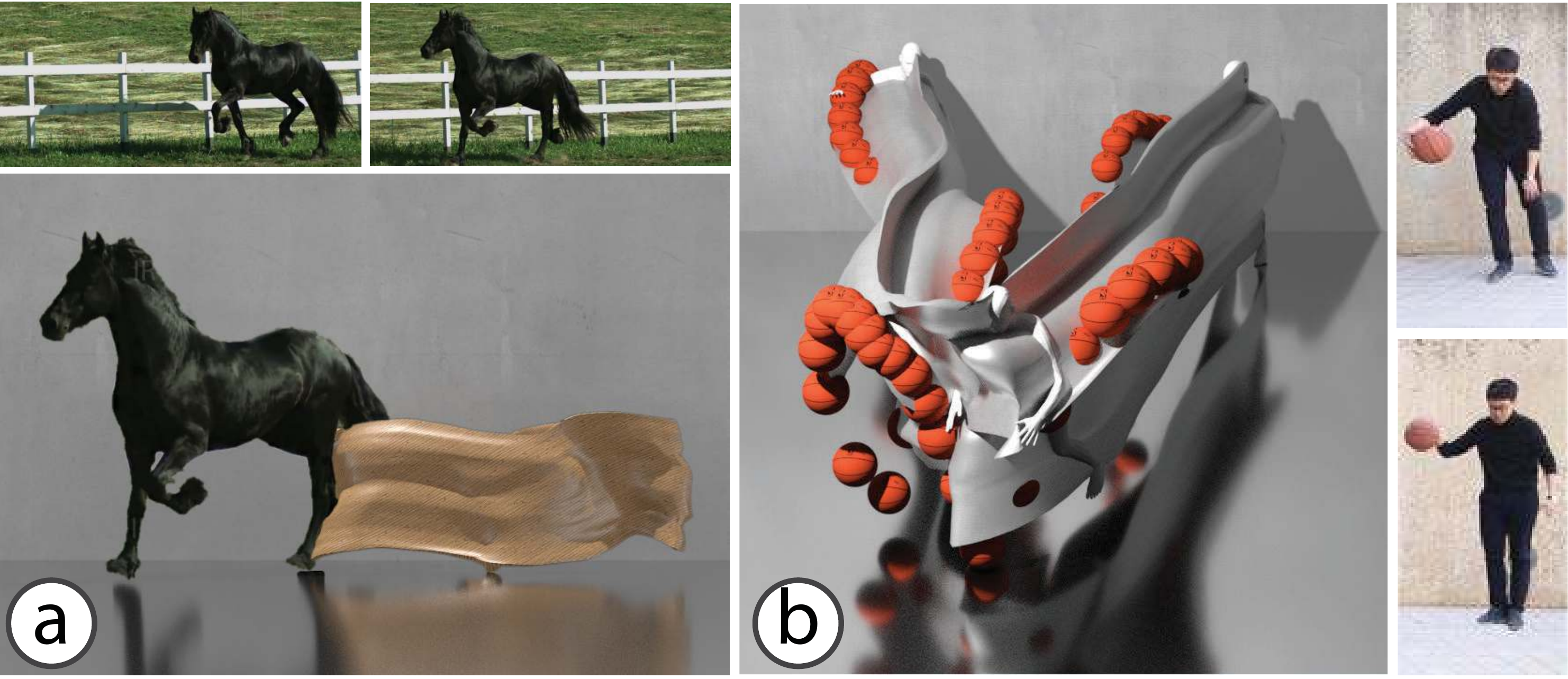}
    % \missingfigure{Figure for horses.}
    % \vspace{-20pt}
    \caption{Non-human \summarynames. We sculpt (a) the leg motion of a horse gait, and (b) the interaction between a basketball and the person dribbling it.}
     \label{fig:non_human}
    %\vspace{-10pt}
\end{figure}

While we have focused on visualizing human motion, our system can also be applied to other objects, as long as they can be reliably represented by a parametric 3D model---an idea that we explore with two examples. \fig{fig:non_human}a shows the \summaryname generated for a running horse, where we visualize its two back legs. To do so, we first estimate the horse's poses across all frames with the per-frame method by Zuffi~\etal~\cite{zuffi20163d}, smooth the estimated poses and translation parameters, and finally apply our method.

In~\fig{fig:non_human}b, we visualize how a basketball interacts in space and time with the person dribbling it. We track the ball in 2D (parameterized by its location and radius), and assign the hand's depth to the ball whenever they are in contact (depth values between two contact points are linearly interpolated). With these depth maps, camera parameters, and ball silhouettes, we insert a 3D ball into the scene.

%We see our work as opening a new direction in visualizing complex 3D motions, such as artistically visualizing the spatial-temporal structure of generic objects and their motions where special skeleton structure is unavailable.  
%When SfM doesn't work, we rely on panorama followed by homography that sometimes stretches some frames, causing inaccurate 2D keypoint detection, which leads to shaking as in the \textit{Lebron slamdunk} sequence, as shown in \fig{fig:limit}a.

\ignore{
We can confess to some of the following technical limitations:
1. Should've used joint heatmap, instead of point estimations\newline
2. Unable to handle partial human bodies, as the pose of the unseen body part is not constrained.\newline
3. Nontrivial to extend to multi-person scenes, as we need to prohibit collision between human meshes.\newline
4. Learning motion priors will be much better than the  pose priors currently used\newline
5. Errors due to depth ambiguities in single-view videos (Do we want to say this as a limitation? Our rendering from the original view is meant to overcome this, so it's not a limitation?)\newline
}
\section{Discussion \& Conclusion}

We presented \modelname, a system that automates the creation of \summarynames, and allows users to interactively explore the visualization and customize various rendering settings. Our system makes motion sculpting accessible to novice users, and requires only a video as input.

% We provided a method for constructing sculptures, based on computer vision techniques and human feedback. We then provided an interactive interface for visualizing these sculptures under different settings, such as by inserting them back into the source videos in a way that respects 3D occlusion relationships. 

\begin{figure}[t]
    \centering
    \includegraphics[width=\linewidth]{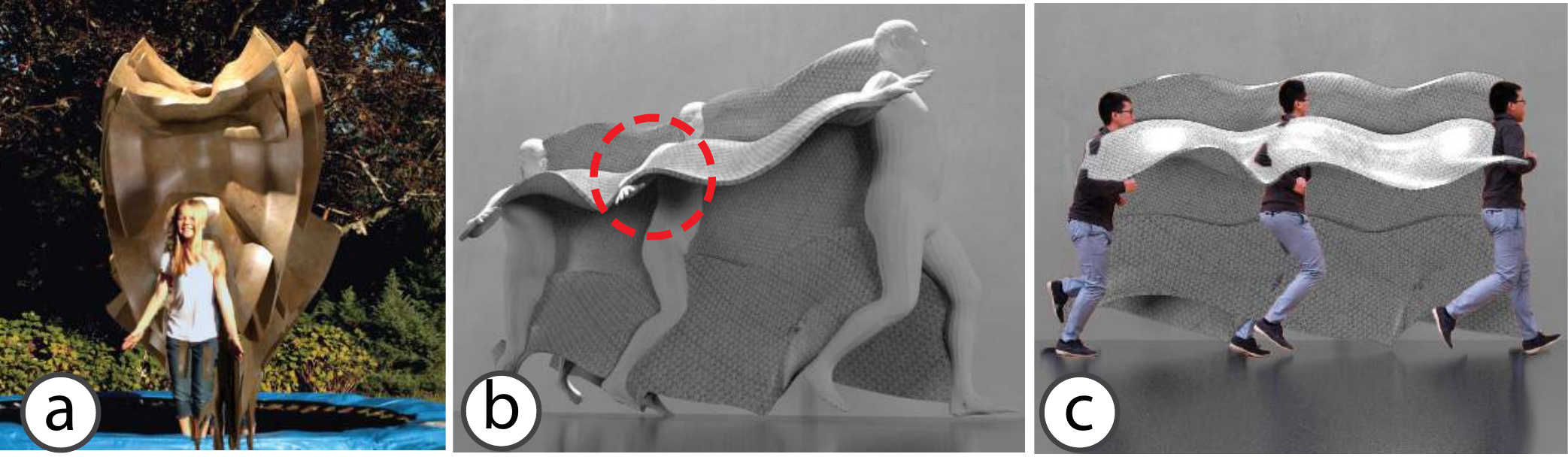}
    % \missingfigure{Limitations}
    % \vspace{-20pt}
    \caption{\change{Limitations. }(a) Cluttered \summaryname due to repeated and spatially localized motion. (b) Inaccurate pose: there are multiple arm poses that satisfy the same 2D projection equally well. (c) Nonetheless, these errors are not noticeable in the original camera view.}
    \label{fig:limit}
    %\vspace{-10pt}
\end{figure}

As for limitations, our \summaryname may look cluttered when the motion is repetitive and spans only a small region (\fig{fig:limit}a). In addition, we rely on high-quality pose estimates, which are sometimes unattainable due to the inherent ambiguity of the 2D-to-3D inverse problem. \fig{fig:limit}b shows such an example\change{: when the person is captured in side profile throughout the video (\fig{fig:limit}c), there are multiple plausible arm poses that satisfy the 2D projection equally well. The red-circled region in \fig{fig:limit}b shows one plausible, but wrong arm pose.} Nevertheless, when our algorithm renders the imperfect sculpture back into the video from its original viewpoint, these errors are no longer noticeable (\fig{fig:limit}c). %Also, while our algorithm works well with large motions that traverse through the space, repetitive motion within a certain spatial volume can lead to significant self-occlusion in its \summaryname. 

%with 78 percent of users rating our visualization as better understandable %when retrieving specific information of 3D motion
We demonstrated our motion sculpting system on diverse videos, revealing complex human motions in sports and dancing. We also demonstrated through user studies that our visualizations facilitate users' understanding of 3D motion. We see two directions opened by this work. The first is in developing artistic tools that allow users to more extensively customize the aesthetics of their renderings, while preserving the interpretability. The second is in rendering \summarynames in other media. In \fig{fig:teaser}d, we showed one example of this---a 3D printed sculpture, and future work could move towards customizing and automating this process.

%With the goal of allowi users to more easily capturehttps://www.sharelatex.com/3588388427wktsnvxzqfcn and communicate complex actions to others, in an aesthetically pleasing form. Future directions therefore include new methods for artistically manipulating the motion sculptures.

%Future directions of our work include reduce rendering time, and developing tools that streamline the process of 3D printing the sculptures.
\section{Acknowledgement}

We thank the anonymous reviewers for their constructive comments. We are grateful to Felice Frankel for her professional advice on aesthetics, and Angjoo Kanazawa for her help in running~\cite{zuffi20163d} on the \emph{Horse} sequence. We thank Kevin Burg for allowing us to use the ballet clips from~\cite{ballet}. We thank Katie Bouman, Vickie Ye, and Zhoutong Zhang for their help with the supplementary video. This work is partially supported by Shell Research, DARPA MediFor, and Facebook Fellowship.

% \input{conclusion}

% REFERENCES FORMAT
% References must be the same font size as other body text.
\newpage
\bibliographystyle{SIGCHI-Reference-Format}
\balance
\bibliography{ref}

\end{document}